\documentclass[12pt]{article}
\catcode`\@=11
\@addtoreset{equation}{section}

\global\arraycolsep=2pt 
\newcommand{\nln}{\nonumber\\}

\newcommand{\Arctan}{\operatorname{Arctan}}
\newcommand{\sn}{\operatorname{sn}}
\newcommand{\cn}{\operatorname{cn}}
\newcommand{\dn}{\operatorname{dn}}
\newcommand{\nd}{\operatorname{nd}}
\newcommand{\sd}{\operatorname{sd}}
\newcommand{\cd}{\operatorname{cd}}

\usepackage{mathrsfs,amsbsy,amssymb,latexsym,amsfonts,amsmath,cite}
\usepackage{graphicx,color}
\usepackage{mathtools}

\allowdisplaybreaks

\begin{document}

\begin{flushright}
\parbox{4cm}
{KUNS-2634 \\ 
DIAS-STP-16-06
\\ \today
} 
\end{flushright}

\vspace*{2cm}

\begin{center}
{\Large \bf 
Melnikov's method in String Theory
}
\vspace*{1.5cm}\\
{\large 
Yuhma Asano$^\dagger$\footnote{E-mail:~yuhma@stp.dias.ie}, 
Hideki Kyono$^\ddagger$\footnote{E-mail:~h\_kyono@gauge.scphys.kyoto-u.ac.jp} 
and Kentaroh Yoshida$^\ddagger$\footnote{E-mail:~kyoshida@gauge.scphys.kyoto-u.ac.jp}} 
\end{center}

\vspace*{0.5cm}

\begin{center}
$^\dagger${\it School of Theoretical Physics, Dublin Institute for Advanced Studies, \\
10 Burlington Road, Dublin 4, Ireland}\\
$^\ddagger${\it Department of Physics, Kyoto University, \\ 
Kitashirakawa Oiwake-cho, Kyoto 606-8502, Japan} 
\end{center}

\vspace{1cm}

\begin{abstract}
Melnikov's method is an analytical way to show the existence of classical chaos generated 
by a Smale horseshoe. It is a powerful technique, though its applicability is somewhat limited. 
In this paper, we present a solution of type IIB supergravity to which Melnikov's method is applicable.
This is a brane-wave type deformation of the AdS$_5\times$S$^5$ background. 
By employing two reduction ans\"atze, we study two types of coupled pendulum-oscillator systems. 
Then the Melnikov function is computed  for each of the systems by following the standard way 
of Holmes and Marsden and the existence of chaos is shown analytically. 
\end{abstract}

\setcounter{footnote}{0}
\setcounter{page}{0}
\thispagestyle{empty}

\newpage

\tableofcontents

\section{Introduction}

The gauge/gravity correspondence is a fascinating topic in the study of string theory. 
A certain class of it preserving a conformal symmetry, in which a string theory is defined 
on an anti-de Sitter (AdS) space and its dual is a conformal field theory (CFT), 
is called the AdS/CFT correspondence \cite{M,GKP,W}. A prototypical example is 
a duality between type IIB string theory on the AdS$_5\times$S$^5$ background and 
the four-dimensional $\mathcal{N}=4$ $SU(N_c)$ super Yang-Mills theory in the large $N_c$ limit. 
Remarkably, the integrability structure exists behind it and hence one can check 
the conjectured relations in a rigorous way by employing various integrability techniques 
(For a big review, see \cite{review}). In particular, the AdS$_5\times$S$^5$ superstring is 
classically integrable in the sense of kinematical integrability \cite{BPR}. 
The integrability exposed in this case is, however, rarely exceptional and 
such a good property is not equipped with general examples of the AdS/CFT (or gauge/gravity) 
correspondence. 

\medskip 

Holographic interpretations in the gauge/gravity correspondence are usually concerned 
with curved string backgrounds and hence classical motions of a string are described 
by non-linear equations. But most of the non-linear equations are not integrable and 
therefore the behavior of classical string solutions should become chaotic. 
More intriguingly, the holographic counterpart of chaotic strings should exist 
on the gauge-theory side as well, but it has not been clarified yet. 
If it has been done, then one could open up a new frontier 
in the study of the gauge/gravity correspondence. 

\medskip 

A well-studied example of non-integrable backgrounds is AdS$_5\times T^{1,1}$ \cite{KW}, 
where $T^{1,1}$ is a five-dimensional Sasaki-Einstein space \cite{Candelas}\footnote{
The coset construction of $T^{1,1}$ has been completed with a supertrace operation \cite{CMY}.}. 
Chaotic string solutions were found in \cite{T11} by computing Poincar\'e sections. 
The chaotic behavior remains even in a near-Penrose limit \cite{T11-ppwave}. 
Similar studies have been done for many backgrounds in the preceding works 
\cite{AdS-soliton,D-brane,complex-beta,NR,gamma}. 

\medskip 

Similarly, chaotic motions of  
D0-branes can also be studied as well as classical strings. 
The D0-brane dynamics is described by a matrix model 
proposed by Banks, Fischler, Shenker and Susskind (BFSS) \cite{BFSS}. 
Chaotic D0-branes in the BFSS matrix model were studied in \cite{chaos-BFSS} 
by following a seminal paper on chaos in a classical Yang-Mills theory \cite{YM}. 
In comparison to the BFSS case, a matrix model on a pp-wave background, 
which was proposed by Berenstein, Maldacena and Nastase (BMN) \cite{BMN}, 
has a strong advantage that there is no flat direction 
and all of the trajectories are definitely bounded. Classical chaos in the BMN matrix model 
was shown in \cite{chaos-BMN} by following \cite{deformed-YM}. 
The chaos at finite temperature was also studied in \cite{Berenstein} in relation to 
the fast scrambler scenario \cite{SS}. 
A very recent work \cite{HMY} investigated chaotic motions of chiral condensates 
in a holographic QCD setup \cite{HMY} and displayed the dependence of Lyapunov exponent 
on the rank of gauge group $N_c$ and 't Hooft coupling $\lambda$\,.  

\medskip

A lot of achievements have been obtained for the chaotic behavior of strings and D-branes 
as introduced above. Motivated by this progress,  we are concerned here with an application 
of Melnikov's method \cite{Melnikov} in the context of the gauge/gravity correspondence. 
This method can show analytically the existence of chaos generated by a Smale horseshoe, 
though its applicability is somewhat restricted. 
In this paper, we will present a string background to which Melnikov's method is applicable. 
This is a brane-wave type deformation of AdS$_5\times$S$^5$ presented in \cite{HY}. 
By employing two reduction ans\"atze, we study two types of coupled 
pendulum-oscillator systems. Then the Melnikov function is computed for each of the systems 
in the standard way of Holmes and Marsden \cite{HolmesMarsden:1982}. 

\medskip 

This paper is organized as follows. 
Section 2 gives a concise review of Melnikov's method and 
provides simple examples. In section 3, we introduce a brane-wave type 
deformation of AdS$_5\times$S$^5$\,, and study two ans\"atze. 
The associated Melnikov function is computed for each of the cases. 
Section  4 is devoted to conclusion and discussion. 
In Appendix \ref{elliptic functions}, the definitions and useful identities of elliptic functions are summarized.
Appendix \ref{apx:calc} explains a detailed computation of Melnikov function for a spinning string ansatz.

\section{Melnikov's method}

In this section, we shall give a brief introduction of Melnikov's method\footnote{
For a concise review, for example, see \cite{Holmes:1990}.}. 
This is an analytical method to argue the existence of classical chaos. 
The key ingredient in this method is the so-called Melnikov function and its simple zero points are related 
to a discrete dynamical system, Smale's horseshoe, which generates a certain class of chaos. 

\medskip 

\begin{figure}[htbp]
 \centering
 \includegraphics[scale=0.35]{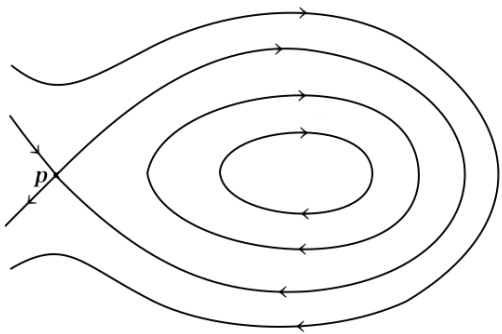}
 \caption{\footnotesize A homoclinic orbit in the $(q_2,p_2)$ phase space 
with a hyperbolic saddle point $p$ and the other two types of orbits. 
The concentric rings inside the homoclinic orbit describe oscillation 
while the open trajectories outside it come from the infinity, surround the homoclinic orbit 
and go away to the infinity again. }
 \label{fig:separatrix}
\end{figure}

Let us concentrate on a continuous dynamical system composed of two particles
which has the Hamiltonian represented by a direct sum like $H=H_1(q_1,p_1)+H_2(q_2,p_2)$\,. 
The four-dimensional phase space is spanned by the coordinates $q_i~(i=1,2)$ and 
the conjugate momenta $p_i$\,. We are concerned with evolution of $q_i$ and $p_i$ in time $\tau$\,. 
It is also supposed that $H_2$ has a homoclinic orbit (Fig.~\ref{fig:separatrix}) 
with a hyperbolic saddle point $p$\,. The existence of the homoclinic orbit is surely crucial 
so as to apply Melnikov's method. 
In other words, this condition somewhat restricts the applicability of this method. 

\medskip

\begin{figure}[htbp]
\begin{center}
\begin{tabular}{cc}
\includegraphics[scale=.21]{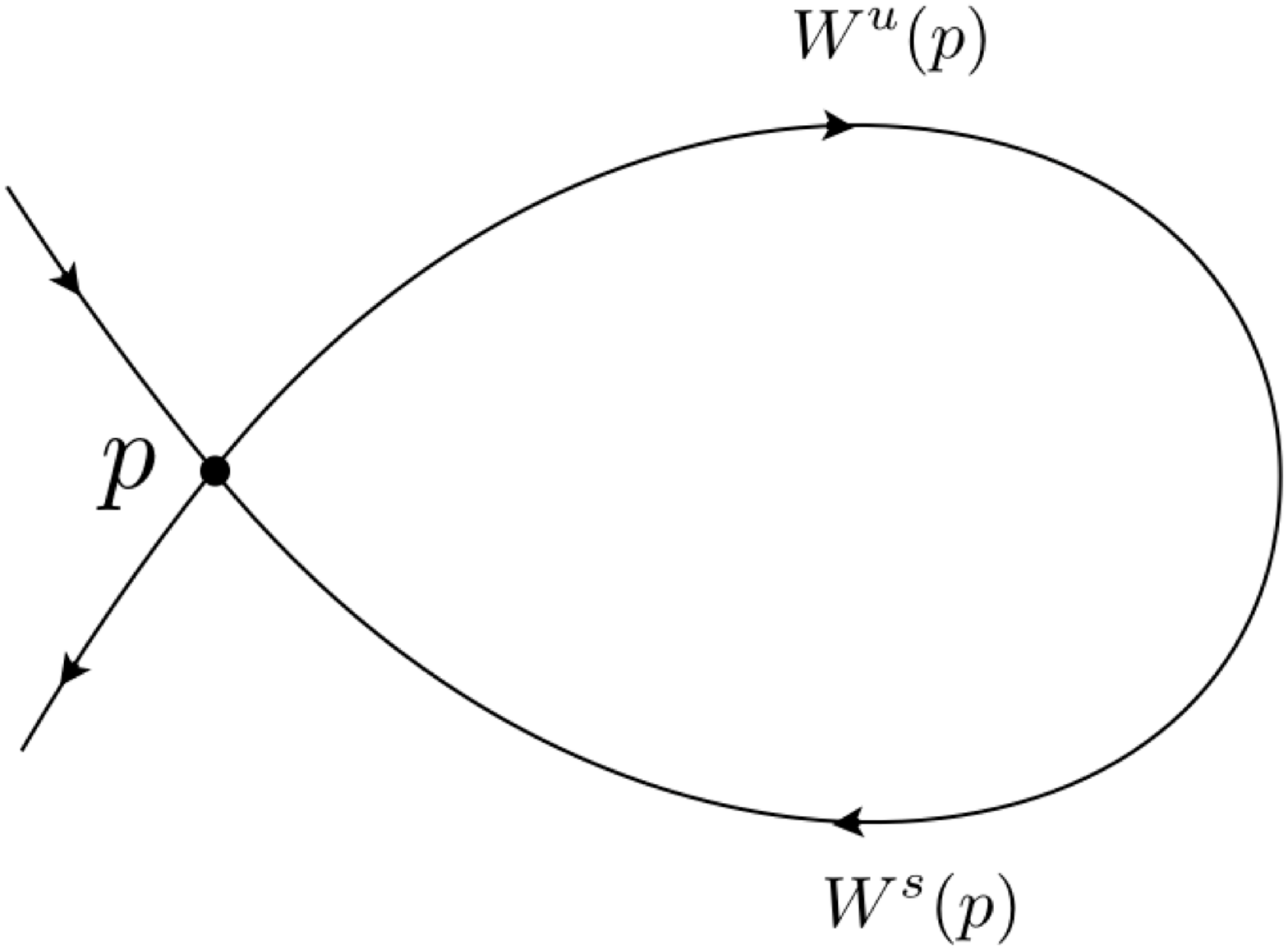} &
\includegraphics[scale=.21]{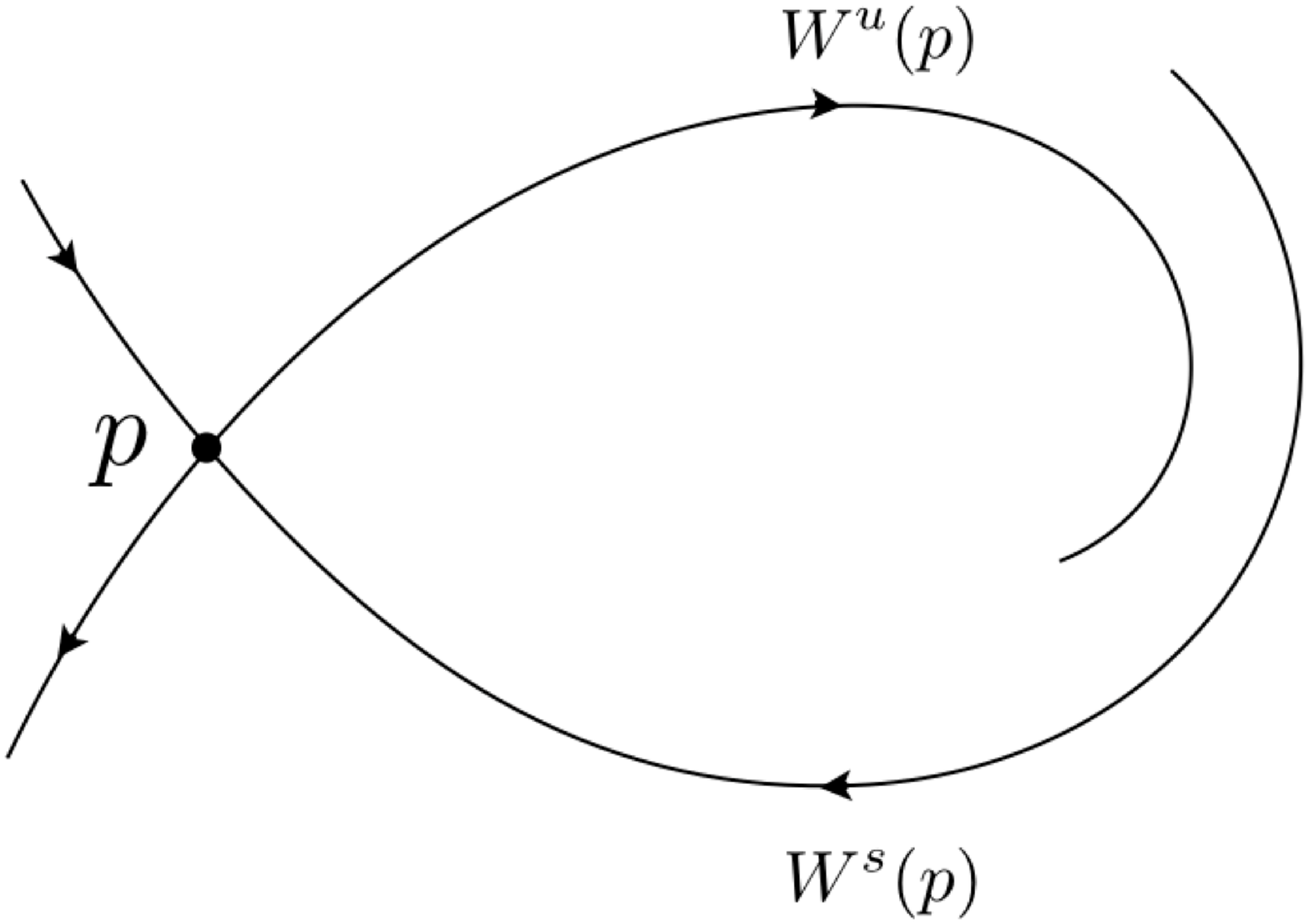} 
\vspace*{-0.7cm} \\
{\footnotesize (a)} &
{\footnotesize (b)}  \\
\end{tabular}
\caption{\label{fig:homoclinic-orbit} \footnotesize In the non-perturbed case (a), 
there is a homoclinic orbit, which is the degeneracy 
of the stable and unstable manifolds around the saddle point $p$\,. 
In the perturbed case (b), the degeneracy is resolved and 
the homoclinic orbit is broken up. Hence the stable and unstable manifolds are now separated.}
\end{center}
\end{figure}

This system is integrable because the Hamiltonian is separable and one can solve the equation 
of motion as a one dimensional Hamiltonian system. 
Then, once a certain small perturbation is added to the integrable system, the integrability is broken
and orbits near the homoclinic one are likely to be chaotic (c.f., KAM theorem \cite{Ko,Ar,Mo}). 
When the perturbation is turned on, the homoclinic orbit opens up in general as in Fig.~\ref{fig:homoclinic-orbit}, 
while the saddle point survives up to a slight deviation of the location 
(i.e., the saddle is not resolved but just moves a little bit). 
Then there still remain a stable manifold $W^s(p)$ and an unstable one $W^u(p)$ which meats at the saddle point.
The stable manifold is defined as the set of points that get infinitely close to the saddle point 
by time evolution. Similarly, the unstable manifold is the set of points getting 
closer to the saddle point by the inverse evolution\footnote{For more rigorous definition, 
for example, see \cite{Wiggins}.}. 
In general, the two manifolds are separated after the system is perturbed, 
while each of them is closed at the saddle point and forms a closed loop called a homoclinic orbit 
(i.e., the degeneracy of the stable and unstable manifolds) in the non-perturbed system. 

\medskip 
 
The idea of Melnikov's method is to argue whether the stable manifold and the unstable manifold 
have a point of transverse intersection, i.e., a transverse homoclinic point. 
When the system is perturbed by a small interaction term $\epsilon H_{\rm{int}}$ 
where $\epsilon$ is a small real parameter, 
the orbits on these manifolds are expressed as perturbative expansions from the 
original ones by an infinitesimal parameter $\epsilon$\,.
Then, one can define the separation between those manifolds, 
which is proportional to the Melnikov function.

\medskip 

\begin{figure}[htbp]
 \centering
 \includegraphics[scale=0.4]{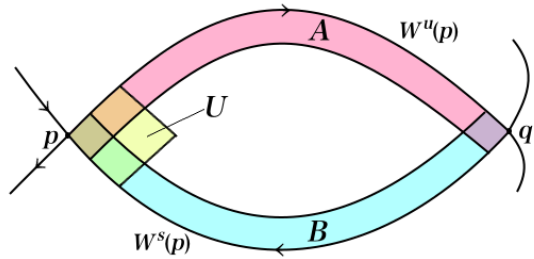}
 \caption{\footnotesize An illustrative explanation of Smale's horseshoe in the phase space. 
This shows orbits that used to be a homoclinic orbit now have an intersection, 
which is a transverse homoclinic point. Due to the existence of the transverse homoclinic point, 
the yellow area $U$ is mapped to the red area $A$\,, 
and thereafter the blue area $B$ is mapped to $U$\,. 
Note that these areas overlap with one another and the colors in the overlapping regions 
are different from the original ones. The entire map is regarded as a horseshoe map. 
A similar argument is applicable to the case with a heteroclinic orbit.}
\label{fig:horseshoe phase}
\end{figure}

\begin{figure}[htbp]
 \centering
 \includegraphics[scale=0.3]{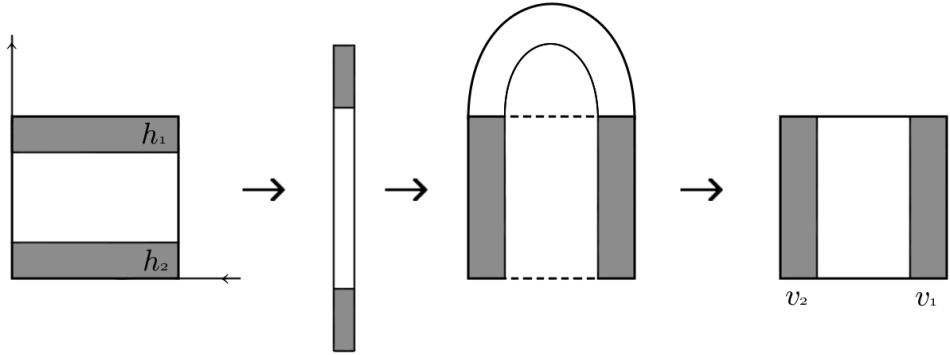}
 \caption{\footnotesize Smale's horseshoe map. 
The horizontal and vertical axes indicate a stable and an unstable manifold, respectively. 
The origin is a hyperbolic saddle point. The horizontal strips $h_1$ and $h_2$ are mapped 
onto the vertical strips $v_1$ and $v_2$\,.
In the end, the initial square is transformed to another square.
These squares correspond to the region $U$ in the phase space.}
 \label{fig:horseshoemap}
\end{figure}

In particular, in the case that at least one transverse homoclinic point exists, 
a certain class of chaos is generated by Smale's horseshoe (see Fig.~\ref{fig:horseshoe phase}).
The saddle point expands small phase-space volume $U$ around the saddle point $p$
into long thin volume $A$ along the unstable manifold to $p$, $W^u(p)$, and
its end tip reaches the stable manifold $W^s(p)$ at a transverse homoclinic point $q$.
Then the area $B$ along the stable manifold is compressed 
into a vicinity around the saddle point, $U$.
This one cycle of process 
transforms $U$ to a U-shaped form, which looks like a horseshoe. 
To be more precise, two horizontal strips $h_1$ and $h_2$ in $U$ 
which move to the intersections of $A$ and $B$ by successive Poincar\'e maps, 
are mapped into two vertical strips $v_1$ and $v_2$ in $U$\,, 
as shown in Fig.~\ref{fig:horseshoemap}.
This is Smale's horseshoe map.
The Smale-Birkhoff homoclinic theorem states that 
this map has a hyperbolic invariant set, which is a direct product of Cantor sets.
By the repetition of this map, a chaotic behavior appears near the saddle point 
and the transverse homoclinic points.
If the Melnikov function has a simple zero point, then the stable manifold 
and the unstable manifold have a transverse intersection 
and the system exhibits chaos due to the above argument. 
Thus the criterion to determine whether chaos of Smale's horseshoe type appears or not 
is boiled down to a simple calculation of the Melnikov function.

\medskip

In order to apply Melnikov's method, it is convenient to 
transform a canonical pair, say $q_1$ and $p_1$\,, into 
the action-angle variables $I$ and $\Theta$\,, 
while the others are kept so as to preserve the homoclinic structure. 
Section \ref{sec:stand-meln-meth} is devoted to the standard review of Melnikov's method. 
Some simple examples are presented in Sec.~\ref{2.2}. 
If the resulting variables do not form a canonical pair,
the Hamiltonian generally cannot be separated into a direct sum, 
and hence Melnikov's method has to be modified. 
Such a case often arises in some studies of a string sigma model with an AdS-like geometry. 
This modification of Melnikov's method will be discussed in Sec.~\ref{sec:spec-meln-meth}.

\subsection{Melnikov's method in 4D systems --the standard case--}\label{sec:stand-meln-meth}

First of all, we will briefly introduce the reduction method of a four-dimensional system with $q_i~(i=1,2)$ 
and the conjugate momenta $p_i$ to a two-dimensional system, as presented in \cite{HolmesMarsden:1982}.  
The non-perturbed Hamiltonian $H^{(0)}$ is supposed to be a direct sum of two classically integrable 
systems like 
\[
H^{(0)}(q_1,p_1,q_2,p_2) = H_1(q_1,p_1) + H_2(q_2,p_2)
\] 
and the interaction between them is turned on as a small perturbation 
$\epsilon\,H_{\rm int}(q_1,p_1,q_2,p_2)$\,, where $\epsilon$ is an infinitesimal real constant. 
That is, the total Hamiltonian $H$ is given by 
\begin{eqnarray}
H(q_1,p_1,q_2,p_2) =  H_1(q_1,p_1) + H_2(q_2,p_2) +\epsilon\,H_{\rm int}(q_1,p_1,q_2,p_2)\,. 
\end{eqnarray}
In the following, we assume that $H_1$ is periodic in $q_1$ and 
$H_2$ has a homoclinic orbit. The existence of the homoclinic orbit is crucial in 
Melnikov's method. 

\subsubsection*{Introducing the action-angle variables}

It is convenient to transform $q_1$ and $p_1$ into the angle-action variables 
$\Theta$ and $I$\,. Then the Hamiltonian $H$ can be rewritten as  
\begin{align}
\label{Hamiltonian1}
 H(\Theta,I,q_2,p_2)=H_1(I)+H_2(q_2,p_2)+\epsilon\, H_{\rm int}(\Theta,I,q_2,p_2).
\end{align}
The Hamilton equations are given by 
\begin{eqnarray}
 && \dot \Theta =\partial_{I}H_1
 +\epsilon\,\partial_{I}H_{\rm int}\,, \qquad  \qquad
\dot I = -\epsilon\,\partial_{\Theta}H_{\rm int}\,,
 \nonumber \\
&& \dot q_2 =\partial_{p_2}H_2
 +\epsilon\,\partial_{p_2}H_{\rm int}\,, \qquad \quad
\dot p_2 =-\partial_{q_2}H_2
 -\epsilon\,\partial_{q_2}H_{\rm int}\,, 
 \label{eom}
\end{eqnarray}
where the symbol ``$\cdot$'' stands for the derivative with respect to $\tau$\,. 
In addition, $H_{\rm int}$ is also supposed to be periodic in $\Theta$\,.

\medskip 

This dynamical system is essentially three-dimensional 
because the energy $E$ is conserved.
Just for convenience, the unperturbed frequency is supposed to be positive, namely 
\[
\Omega(I)=\partial_IH_1(I) >0\,. 
\] 
When $\epsilon$ is sufficiently small, $\Theta$ becomes a monotonically increasing function 
from the first equation in (\ref{eom}).
Then, thanks to the $\tau$-independence in the dynamical system, 
one can set $\Theta$ as a new time for the system by deleting $\tau$\,.

\subsubsection*{Reduction method}

The next task is to reduce the remaining degree of freedom $I$\,.
By solving the energy conservation $E=H(\Theta,I,q_2,p_2)$\,, $I$ is represented by 
\begin{align}
 &I=L^{(0)}(q_2,p_2)+\epsilon\, L^{(1)}(q_2,p_2,\Theta)+O(\epsilon^2)\,. 
\end{align}
Here $L^{(0)}$ and $L^{(1)}$ depend on $E$ and hence these are also expressed as 
\begin{eqnarray}
\label{L0L1}
L^{(0)}(q_2,p_2)&=& H^{-1}_1 
\left(E-H_2(q_2,p_2)\right)\,,\nln
L^{(1)}(q_2,p_2,\Theta)&=&
-\frac{H_{\rm int}\left(
\Theta,L^{(0)}(q_2,p_2),q_2,p_2
\right)
}{\Omega \left(L^{(0)}(q_2,p_2)\right)}\,.
\end{eqnarray}
Then one can obtain a two-dimensional system which is composed of $q_2$ and $p_2$ with ``time'' $\Theta$:
\begin{align}
 q_2'&=P^{(0)}+\epsilon\, P^{(1)}+O(\epsilon^2)\,,
 \qquad
 p_2'=F^{(0)}+\epsilon\, F^{(1)}+O(\epsilon^2)\,.
 \label{eq:q,p}
\end{align}
Here the prime stands for the derivative with respect to $\Theta$\,, and $P^{(0)}$\,, $P^{(1)}$\,, 
$F^{(0)}$ and $F^{(1)}$ are given by, respectively,   
\begin{align}
 P^{(0)} & =\frac{1}{\Omega(L^{(0)}(q_2,p_2))}\,\partial_{p_2}H_2(q_2,p_2)\,,
 \nonumber \\
 P^{(1)} & =\frac{\partial_{p_2}H_{\rm int}(\Theta,L^{(0)},q_2,p_2)}{\Omega(L^{(0)}(q_2,p_2))}
 -\biggl[
\partial_{I}H_{\rm int}(\Theta,L^{(0)},q_2,p_2)\,, 
 \nonumber \\
 &\hspace{55mm}
 +\Omega'(L^{(0)}(q_2,p_2))\, L^{(1)}(q_2,p_2,\Theta)
 \biggr]
  \frac{\partial_{p_2}H_2(q_2,p_2)}{\Omega(L^{(0)}(q_2,p_2))^2}\,, 
 \nonumber \\
 F^{(0)} & =-\frac{1}{\Omega(L^{(0)}(q_2,p_2))}\,\partial_{q_2}H_2(q_2,p_2)\,,
 \nonumber \\
 F^{(1)} & =-\frac{\partial_{q_2}H_{\rm int}(\Theta,L^{(0)},q_2,p_2)}{\Omega(L^{(0)}(q_2,p_2))}
 +\biggl[
\partial_{I}H_{\rm int}(\Theta,L^{(0)},q_2,p_2)\,, 
 \nonumber \\
 &\hspace{55mm}
 +\Omega'(L^{(0)}(q_2,p_2))\, L^{(1)}(q_2,p_2,\Theta)
 \biggr]
  \frac{\partial_{q_2}H_2(q_2,p_2)}{\Omega(L^{(0)}(q_2,p_2))^2}\,.
\end{align}
Note that in the above computation we have used the expressions in \eqref{L0L1} 
and the following relations:
\begin{eqnarray}
 && q'_2=\frac{\dot{q}_2}{\dot{\Theta}}\,,
  \qquad p'_2=\frac{\dot{p}_2}{\dot{\Theta}}\,.
\end{eqnarray}

\subsubsection*{Melnikov's method}

We are now ready to introduce Melnikov's method. 

\medskip

To explain the role of the Melnikov function in this method, 
we employ the Poincar\'e map at $\Theta=\Theta_0$ with a perturbation parameter $\epsilon$
by $P^{\Theta_0}_{\epsilon}$\,, where $\Theta_0$ is a value modulo the period.
By this Poincar\'e map, a point on a orbit at $\Theta_0$ is mapped 
to another on the orbit at $\Theta_0$, modulo the period.
Let us consider a small vicinity $R$ 
around the transverse homoclinic point on the orbit.
This can be mapped to a neighborhood of the saddle point $U$
by finite numbers of forward or backward actions of $P^{\Theta_0}_{\epsilon}$\,.
Since these forward and backward actions can be connected at $U$ and $R$\,, 
there is an integer $N$ 
where $(P^{\Theta_0}_{\epsilon})^{N}$ maps 
a point in $U$ to another in $U$ itself. 
This is Smale's horseshoe. 

\medskip 

Therefore, what we have to do here is to show 
the existence of the transverse homoclinic point.
To verify the condition to ensure its existence,
let us calculate a separation between the stable and unstable manifolds 
passing through the saddle point, at $\Theta=\Theta_0$\,.
One can express a solution of \eqref{eq:q,p} on the stable manifold as
\begin{align}
 q_2^{s}(\Theta,\Theta_0)
 &=q_2^{(0)}(\Theta-\Theta_0)+\epsilon\, q_2^{s(1)}(\Theta,\Theta_0)+O(\epsilon^2)\,,
 \nonumber \\
 p_2^{s}(\Theta,\Theta_0)
 &=p_2^{(0)}(\Theta-\Theta_0)+\epsilon\, p_2^{s(1)}(\Theta,\Theta_0)+O(\epsilon^2)
\end{align}
for $\Theta_0 \leq \Theta < \infty$, and on the unstable manifold as
\begin{align}
 q_2^{u}(\Theta,\Theta_0)
 &=q_2^{(0)}(\Theta-\Theta_0)+\epsilon\, q_2^{u(1)}(\Theta,\Theta_0)+O(\epsilon^2)\,,
 \nonumber \\
 p_2^{u}(\Theta,\Theta_0)
 &=p_2^{(0)}(\Theta-\Theta_0)+\epsilon\, p_2^{u(1)}(\Theta,\Theta_0)+O(\epsilon^2)
\end{align}
for $-\infty < \Theta \leq \Theta_0$\,.
Here the variables $q_{2}^{(0)}$ and $p_{2}^{(0)}$ describe 
a separatrix solution to \eqref{eq:q,p} at $\epsilon=0$\,.
Also, let $q_2^{i(1)}$ and $p_2^{i(1)}$ lie on the normal to 
the unperturbed homoclinic orbit at $\Theta_0$ 
for clarity. Then the separation at $\Theta_0$ can be defined as
\begin{align}
 d(\Theta_0)
 & \equiv \frac{(-F^{(0)}, P^{(0)})\cdot \{(q_2^{u}(\Theta_0,\Theta_0),p_2^{u}(\Theta_0,\Theta_0))-(q_2^{s}(\Theta_0,\Theta_0),p_2^{s}(\Theta_0,\Theta_0))\}}
 {\sqrt{P^{(0)\; 2}+F^{(0)\; 2}}}
 \nonumber \\
 &=\epsilon\,\frac{P^{(0)}(p_2^{u(1)}(\Theta_0,\Theta_0)-p_2^{s(1)}(\Theta_0,\Theta_0))
 -F^{(0)}(q_2^{u(1)}(\Theta_0,\Theta_0)-q_2^{s(1)}(\Theta_0,\Theta_0))}
 {\sqrt{P^{(0)\; 2}+F^{(0)\; 2}}} \nonumber \\ 
& \quad +O(\epsilon^2) \,,
\label{distance-d}
\end{align}
where $P^{(0)}$ and $F^{(0)}$ are functionals of $q_2^{(0)}(\Theta-\Theta_0)$ 
and $p_2^{(0)}(\Theta-\Theta_0)$\,.
This is the projection to a vector $(-F^{(0)},P^{(0)})$ which is normal to the original homoclinic orbit.
i.e., this vector is orthogonal to $(q^{(0)}_2{}', p^{(0)}_2{}')=(P^{(0)},F^{(0)})$\,. 
\begin{figure}[htbp]
 \centering
 \includegraphics[scale=0.3]{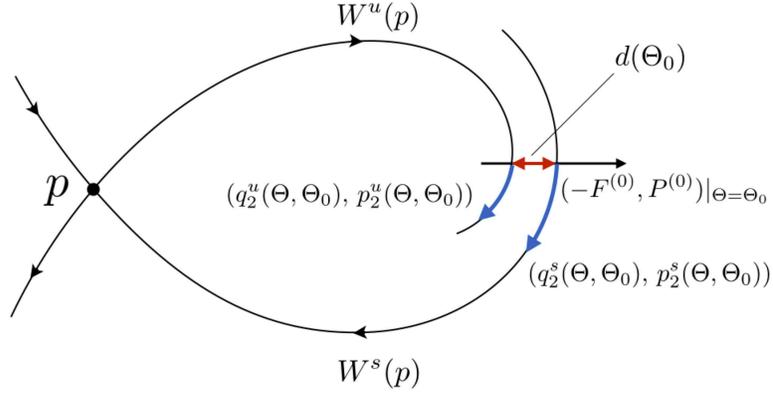}
 \vspace*{-1.2cm}
 \caption{\footnotesize $d(\Theta_0)$ is a distance between the stable and unstable manifolds.}
 \label{fig:melnikov-distance}
\end{figure}

\medskip 

For convenience, let us introduce $\Delta^i(\Theta,\Theta_0)$ defined as 
\[
\Delta^{i}(\Theta,\Theta_0) \equiv 
P^{(0)} p_2^{i(1)}(\Theta,\Theta_0)-F^{(0)} q_2^{i(1)}(\Theta,\Theta_0)\,, 
\]
where $i=s$ and $u$\,. Then the derivative of $\Delta^i(\Theta,\Theta_0)$ is represented by 
\begin{align}
 \frac{d}{d\Theta}\Delta^i(\Theta,\Theta_0)
 =P^{(0)}F^{(1)}-F^{(0)}P^{(1)} \,.
\end{align}
Here $P^{(1)}$ and $F^{(1)}$ are functionals of 
$q_2^{(0)}(\Theta-\Theta_0)$\,, $p_2^{(0)}(\Theta-\Theta_0)$\,, 
and explicit $\Theta$\,. Note that we have utilized the following relations: 
\begin{eqnarray}
\frac{\partial P^{(0)}}{\partial q_2} &=& -\frac{\partial F^{(0)}}{\partial p_2}\,, \nonumber \\  
q_2^{i(1)\prime} &=& \frac{\partial P^{(0)}}{\partial q_2}q_2^{i(1)}
+\frac{\partial P^{(0)}}{\partial p_2}p_2^{i(1)}+P^{(1)}\,, \nonumber \\ 
p_2^{i(1)\prime} &=&\frac{\partial F^{(0)}}{\partial q_2}q_2^{i(1)}
+\frac{\partial F^{(0)}}{\partial p_2}p_2^{i(1)}+F^{(1)}\,.
\end{eqnarray}

\medskip 

By employing the numerator in (\ref{distance-d})\,, let us define the Melnikov function as 
\begin{align}
 M(\Theta_0)
 &\equiv P^{(0)}\cdot (p_2^{u(1)}(\Theta_0,\Theta_0)-p_2^{s(1)}(\Theta_0,\Theta_0))
 -F^{(0)}\cdot (q_2^{u(1)}(\Theta_0,\Theta_0)-q_2^{s(1)}(\Theta_0,\Theta_0))\,.
 \label{Melnikov fn 1D}
\end{align}
Hence, as far as the leading order is concerned, 
the separation $d(\Theta_0)$ is proportional to the Melnikov function like 
\begin{align}
 d(\Theta_0)
 &=\epsilon\frac{M(\Theta_0)}
 {\sqrt{P^{(0)\; 2}+F^{(0)\; 2}}}
 +O(\epsilon^2) \,.
\end{align}
By using $\Delta^i$\,, the Melnikov function (\ref{Melnikov fn 1D}) can be rewritten as
\begin{eqnarray}
M(\Theta_0)&=&\Delta^u(\Theta_0,\Theta_0)-\Delta^s(\Theta_0,\Theta_0)\nonumber\\
&=&\int_{-\infty}^{\Theta_0}\!\! d\Theta \frac{d}{d\Theta} \Delta^u(\Theta,\Theta_0)
+\int^\infty_{\Theta_0}\!\! d\Theta \frac{d}{d\Theta} \Delta^s(\Theta,\Theta_0)\nonumber\\
&=&\int^\infty_{-\infty}\!\! d\Theta (P^{(0)}F^{(1)}-F^{(0)}P^{(1)}) 
(q_{2}^{(0)}(\Theta-\Theta_0),p_{2}^{(0)}(\Theta-\Theta_0),\Theta)\,. 
\label{Mel-t}
\end{eqnarray}
Here we have used the following relations:
\begin{eqnarray}
\Delta^s(\infty,\Theta_0)=\Delta^u(-\infty,\Theta_0)=0\,. \nonumber 
\end{eqnarray}
Note that near the saddle point, the derivatives of $q_2$ and $p_2$ are equal to zero.
Hence $P^{(0)}|_{\Theta=\pm\infty}=F^{(0)}|_{\Theta=\pm\infty}=0$ and the above equations 
follow\footnote{Note that the definition of Melnikov function 
follows from the existence of just two ingredients: a set of two equations of motion 
as in (\ref{eq:q,p}) and a non-perturbed homoclinic solution.}.

\medskip 

Now the argument of the Melnikov function (\ref{Mel-t}) is given by $\Theta_0$\,. 
It is more useful to rewrite the Melnikov function as a function of $\tau$\,, rather than $\Theta$\,. 
Let us define $I^{(0)}$ as the initial value of $I$\,. Then the time $\tau$ can be expressed as 
\[
\tau=\frac{\Theta}{\Omega(I^{(0)})} +O(\epsilon)\,.
\] 
Finally, the Melnikov function is represented by  
\begin{align}
 M(\tau_0) &=\int_{-\infty}^\infty\!\! d\tau\, 
 \frac{1}{\Omega(I^{(0)})}
 \{ H_2, H_{\rm int} \}_2 (q_{2}^{(0)}(\tau-\tau_0),p_{2}^{(0)}(\tau-\tau_0),I^{(0)},\tau)
 +O(\epsilon)\,,
 \label{Melnikov fn reduced}
\end{align}
where $q_{2}^{(0)}$ and $p_{2}^{(0)}$ have been redefined as functions of $\tau$\,,  
and $\{ \; ,\; \}_2$ is the Poisson bracket defined with $q_2$ and $p_2$ like
\begin{align}
\{ f,g \}_2 \equiv \partial_{q_2}f\,\partial_{p_2}g-\partial_{p_2}f\,\partial_{q_2}g\,.
\end{align}
According to Melnikov's method,
the existence of simple zeros in $M(\tau_0)$ means that
this system has a horseshoe for sufficiently small $\epsilon$
lying near the homoclinic orbit of $(q_2,p_2)$ on the energy surface $H=E$\,.

\subsection{Examples of computing Melnikov functions \label{2.2}}

It would be helpful to see examples of computing the Melnikov functions. 
Let us show here two examples: 1) a dynamical system
composed of a coordinate variable $q$ and its conjugate momentum $p$ 
with a time-dependent perturbation, 2) a four-dimensional dynamical system  
composed of two coordinate variables $q_1$\,, $q_2$ and their conjugate momenta $p_1$\,, $p_2$ 
with a time-independent perturbation. These examples are discussed in detail 
in \cite{Holmes:1990} and \cite{HolmesMarsden:1982}, respectively. 

\medskip 

More precisely speaking, the first example is a three-dimensional dynamical 
system\footnote{When the Hamiltonian depends 
explicitly on time, the time is regarded as an additional direction of the phase space.}. 
This case has not been covered in the previous subsection, but the computation scheme is almost the same. 
A central difference is that the reduction process is not necessary. 
It would really be instructive to study this case and to capture the essential idea of the Melnikov function. 

\subsubsection*{1) a three-dimensional dynamical system}
 
We consider a pendulum system with a periodic external force 
described by the Hamiltonian, 
\begin{align}
H=\frac{1}{2}p^2-\cos q-\epsilon\, q\cos\tau \,,
\end{align}
and it depends on time $\tau$ explicitly. Then the Hamilton equations are given by
\begin{eqnarray}
 \dot q=p\,,  \qquad \dot p=-\sin q+\epsilon \cos \tau \,.
 \label{example1}
\end{eqnarray}
Here $\epsilon$ is an infinitesimal constant. 
Note that these equations can be regarded as those of reduced systems 
like (\ref{eq:q,p}) by replacing $\tau$ with $\Theta$\,. 
This point is also concerned with the footnote 5.

\medskip 

When the perturbation is turned off (i.e., $\epsilon=0$) and the energy $E$ is set to 1, 
this system exhibits a separatrix solution given by 
\begin{align}
 q^{(0)}(\tau)= \pm 2\Arctan (\sinh\tau)\,,
 \qquad
 p^{(0)}(\tau)=\pm \frac{2}{\cosh\tau}\,. 
 \label{example1-sol}
\end{align}
The pendulum simply oscillates around the stable fixed point $q=0$ below $E=1$\,, 
while the pendulum rotates all the way above $E=1$\,.

\medskip 

Then by substituting the reduced equations (\ref{example1}) and the homoclinic solution 
(\ref{example1-sol}) into the formula (\ref{Melnikov fn reduced})\,, 
the Melnikov function can be computed as  
\begin{align}
 M(\tau_0)
 &=\int_{-\infty}^\infty\!\! d\tau\; p^{(0)}(\tau-\tau_0)\,\cos\tau
 \nonumber \\
 &=\int_{-\infty}^\infty\!\! d\tau\, \frac{2\cos\tau}{\cosh(\tau-\tau_0)}
 =\frac{2\pi\cos\tau_0}{\cosh\left(\frac{\pi}{2}\right)}\,.
\end{align}
It has nontrivial simple zeros at $\tau_0=\pi/2+\pi l~(l\in \mathbb{Z})$\,.
Hence, when $E$ is close to $1$\,, the pendulum behaves in a weird manner
--- it switches oscillation and rotation randomly. This is nothing but the origin of 
chaotic motion. Melnikov's method tells us that
this chaotic motion is generated by Smale's horseshoe.

\subsubsection*{2) a four-dimensional dynamical system}

As the second example, let us look at a simple pendulum-oscillator system 
with a small perturbation. This is a four-dimensional dynamical system composed of 
$q_1$\,, $q_2$ and the conjugate momenta $p_1$ and $p_2$\,.
By taking $\epsilon\, H_{\rm int}=\epsilon\,(q_2-q_1)^2/2$ as a perturbation term, 
the perturbed Hamiltonian is given by 
\begin{align}
 H
 =\frac{1}{2}\left(
 p_1^2+\omega^2q_1^2
 \right)
 +\frac{p_2^2}{2}-\cos q_2
 +\frac{\epsilon}{2}(q_2-q_1)^2\,.
\end{align}
Here $\epsilon$ is an infinitesimal parameter again and 
$\omega$ is a frequency of the oscillator. 

\medskip 

When the perturbation is turned off (i.e., $\epsilon=0$), 
the system is completely separable: 
the one is a simple oscillator and the other is a pendulum.
The variables of the oscillator $q_1$ and $p_1$ can be transformed into 
the action and angle variables, $I$ and $\Theta$\,. 
Then the Hamiltonian can be rewritten as 
\begin{align}
 H
 =\omega I
 +\frac{p_2^2}{2}-\cos q_2
 +\frac{\epsilon}{2}\left(
 q_2-\sqrt{\frac{2I}{\omega}}\sin\Theta
 \right)^2\,.
\end{align}
For the unperturbed system, 
the oscillation can be written as $\Theta=\omega\tau$,
and the separatrix solution of the unperturbed subsystem with $(q_2,p_2)$  
is the same as in the previous.

\medskip 

When the perturbation is turned on,
the pendulum and oscillator systems begin to interact each other weakly, 
and the pendulum feels the motion of the oscillator through the interaction term.
This situation is quite similar to the dynamical system \eqref{example1}.
Thus, as in Sec.\ 3.1, the reduction method leads to a two-dimensional dynamical system 
to which Melnikov's method can be applied. 

\medskip 

In the present case, one can straightforwardly employ the formula \eqref{Melnikov fn reduced}. 
By putting $H_{\rm int}$ and $H_2=\frac{p_2^2}{2}-\cos q_2$ into \eqref{Melnikov fn reduced},
the Melnikov function can be evaluated as  
\begin{align}
 M(\tau_0)
 &=\frac{1}{\omega}\int_{-\infty}^\infty\!\! d\tau \;
 p_2^{(0)}(\tau-\tau_0) \left[
 \sqrt{\frac{2I^{(0)}}{\omega}}\sin\omega \tau-q_2^{(0)}(\tau-\tau_0)
 \right]
 \nonumber \\
 &=\frac{1}{\omega}\int_{-\infty}^\infty\!\! d\tau \;
 \sqrt{\frac{2I^{(0)}}{\omega}}\, 
 \frac{2\sin\omega \tau}{\cosh(\tau-\tau_0)}
 =\sqrt{\frac{2I^{(0)}}{\omega}}\, 
 \frac{2\pi\sin\omega \tau_0}{\cosh(\frac{\pi\omega}{2})}\,. 
\label{ex2}
\end{align}
It is easy to see that the Melnikov function (\ref{ex2})
has simple zeros at $\tau_0=\pi l/\omega~(l \in \mathbb{Z})$\,.
Hence this system also exhibits chaotic motions generated by Smale's horseshoe.

\subsection{Melnikov's method --a non-direct sum case--}
\label{sec:spec-meln-meth}

So far, we have considered Melnikov's method in the standard case. 
For later purpose, however, we need to generalize it slightly. 
The method we discuss here is a generalization of 
the reduction method in \cite{HolmesMarsden:1982}. 

\medskip 

In this subsection, we are interested in the Melnikov function in more intricate dynamical systems. 
In general, one of $H_i$'s may depend on the other one. Namely, we have in our mind 
the Hamiltonian of the following type 
\begin{align}
 H(q_i,p_i,\tau)=H_1(q_1,p_1,H_2(q_2,p_2))+\epsilon\, H_{\rm int}(q_i,p_i,\tau)\,,
\end{align}
where the entire $H_{\rm int}$ is supposed to be periodic in $\tau$ hereafter.
Then a system composed of $q_1$ and $p_1$ continues to feel a potential due to the energy of $H_2$\,,  
even though the perturbation is turned off, i.e., $\epsilon=0$\,. 
In such a system, one can formally change $q_1$ and $p_1$ to $I$ and $\Theta$ (keeping $q_2$ and $p_2$ unchanged)\,. 
But $I$ and $\Theta$ fail to be a canonical pair. 

\medskip 

The Hamilton equations are given by  
\begin{align}
 \dot q_1&=\partial_{p_1}H_1
 +\epsilon\,\partial_{p_1}H_{\rm int}\,,
 \nonumber \\
 \dot p_1&=-\partial_{q_1}H_1
 -\epsilon\,\partial_{q_1}H_{\rm int}\,,
 \nonumber \\
 \dot q_2&=\partial_{H_2}H_1\partial_{p_2}H_2
 +\epsilon\,\partial_{p_2}H_{\rm int}\,,
 \nonumber \\
 \dot p_2&=-\partial_{H_2}H_1\partial_{q_2}H_2
 -\epsilon\,\partial_{q_2}H_{\rm int}\,. 
\end{align}
Now we have to take account of an additional equation coming from the explicit time-dependence of the Hamiltonian:
\begin{align}
 \dot H(q_i,p_i,\tau)
 =\epsilon\, \partial_\tau H_{\rm int}(q_i,p_i,\tau)\,.
\end{align}

\medskip 

For a moment, let us consider the case with $\epsilon=0$\,. Then the variables of the system become separable. 
When we concentrate on $H_1$ by setting $H_2$ as a constant $h$\,, 
the action-angle variables can be written as
\begin{align}
 I=I(q_1,p_1,h)\,,
 \qquad
 \Theta=\Theta(q_1,p_1,h)\,. \label{2.29}
\end{align}
Here $I$ and $\Theta$ satisfy
\begin{align}
 \partial_{q_1}I\partial_{p_1}H_{1}-\partial_{p_1}I\partial_{q_1}H_{1}=0\,,
 \qquad
 \partial_{q_1}\Theta\partial_{p_1}H_{1}-\partial_{p_1}\Theta\partial_{q_1}H_{1}=\Omega(h)\,, 
 \label{2.30}
\end{align}
and inversely, 
\begin{eqnarray}
q_1=q_1(I,\Theta,h)\,, \qquad p_1=p_1(I,\Theta,h)\,. 
\label{2.31}
\end{eqnarray}

\medskip 

Let us return to the interacting case ($\epsilon \neq 0$). 
By using the relations (\ref{2.29}) and (\ref{2.31}), $I$ and $\Theta$ can be used as new variables, 
instead of $q_1$ and $p_1$\,, even if $\epsilon$ is turned on.
Although the new variables are not canonical,
one can still track their time evolution through
\begin{align}
 \dot I
 &=\epsilon\, (
 \partial_{p_1}H_{\rm int}\,\partial_{q_1}I-\partial_{q_1}H_{\rm int}\,\partial_{p_1}I
 +(\partial_{p_2}H_{\rm int}\,\partial_{q_2}H_2-\partial_{q_2}H_{\rm int}\,\partial_{p_2}H_2)\,\partial_{H_2}I
 )\,,
 \nonumber \\
 \dot \Theta
 &=\Omega+\epsilon\, (
 \partial_{p_1}H_{\rm int}\,\partial_{q_1}\Theta-\partial_{q_1}H_{\rm int}\,\partial_{p_1}\Theta
 +(\partial_{p_2}H_{\rm int}\,\partial_{q_2}H_2-\partial_{q_2}H_{\rm int}\,\partial_{p_2}H_2)\,\partial_{H_2}\Theta
 ) 
 \nonumber \\
 &=\Omega+\epsilon\, \Omega^{(1)}(q_i,p_i,\tau)\,. \label{rational}
\end{align}
In general, $\Omega$ depends on $H_2$ and is assumed to be a positive constant. 
Furthermore, suppose that a ratio of $\Omega$ and the period of $H_{\rm int}$ is rational 
for a reason explained later.
Then, $\tau$ can be expressed as 
\[
\tau= \frac{\Theta}{\Omega} + O(\epsilon) 
\]
and $H_{\rm int}$ is periodic in $\Theta$ to the leading order in $\epsilon$\,.
Since the $\tau$-dependence appears only in $H_{\rm int}$\,, $\tau$ can be replaced by $\Theta/\Omega$ 
as far as the quantities are computed up to the first order of $\epsilon$\,.
When $\epsilon$ is sufficiently small, $\dot\Theta$ becomes positive;
hence $\Theta$ may be regarded as a new time coordinate for the system again. 

\medskip 

Now it is a turn to reduce the degrees of freedom from $q_i,p_i,\tau,H$ to $q_2,p_2,H$\,. 

\medskip 

Let us first remove $\tau$ in $H_{\rm int}$ by $\tau=\Theta/\Omega$ and the derivative 
with respect to $\tau$ by $\dot\Theta\frac{d}{d\Theta}$\,.
The next step is to rewrite functions of $q_1$ and $p_1$ in terms of $I$, $\Theta$ and $H_2$\,. 
Namely, the concerned quantities are rewritten as follows: 
\begin{align}
 &G_{1}(H_2(q_2,p_2),I,\Theta) := \partial_{H_2}H_1(q_1,p_1,H_2(q_2,p_2))\,,
 \nonumber \\
 &\tilde H_{\rm int}(q_2,p_2,I,\Theta) := H_{\rm int}(q_i,p_i,\tau)+O(\epsilon)\,,
 \nonumber \\
 &G_{\mathrm{int},p}(q_2,p_2,I,\Theta) := \partial_{p_2}H_{\rm int}(q_i,p_i,\tau)+O(\epsilon)\,,
 \nonumber \\
 &G_{\mathrm{int},q}(q_2,p_2,I,\Theta) := \partial_{q_2}H_{\rm int}(q_i,p_i,\tau)+O(\epsilon)\,,
 \nonumber \\
 &G_{\mathrm{int},\tau}(q_2,p_2,I,\Theta) := \partial_\tau H_{\rm int}(q_i,p_i,\tau)+O(\epsilon)\,,
 \nonumber \\
 &\tilde\Omega (q_2,p_2,I,\Theta) := \Omega^{(1)}(q_i,p_i,\tau)\,.
\end{align}
Finally, one can eliminate $I$ through the relation 
\begin{align}
 &I=L^{(0)}(H_2(q_2,p_2),H)+\epsilon\, L^{(1)}(q_2,p_2,H,\Theta)+O(\epsilon^2)\,,
\end{align}
with the help of the following implicit function:
\begin{align}
 H=H_1(I,H_2)+\epsilon\, \tilde H_{\rm int}(q_2,p_2,I,\Theta)+O(\epsilon^2)\,. 
\end{align}
The resulting expressions are given by 
\begin{align}
 q_2'&=P^{(0)}+\epsilon\, P^{(1)}+O(\epsilon^2)\,,
 \qquad
 p_2'=F^{(0)}+\epsilon\, F^{(1)}+O(\epsilon^2)\,,
 \nonumber \\
 H'&=\epsilon\, \frac{G_{\mathrm{int},\tau}(q_2,p_2,L^{(0)},\Theta)}{\Omega}+O(\epsilon^2)\,,
 \label{eq:q,p,H}
\end{align}
where we have introduced the following quantities: 
\begin{align}
 P^{(0)}
 &=\frac{G_{1}(H_2,L^{(0)},\Theta)}{\Omega}\,\partial_{p_2}H_2(q_2,p_2)\,,
 \nonumber \\
 P^{(1)}
 &=\frac{G_{\mathrm{int},p}(q_2,p_2,L^{(0)},\Theta)}{\Omega}
 -\biggl[ 
 \frac{G_{1}(H_2,L^{(0)},\Theta)}{\Omega^2}\, \tilde\Omega(q_2,p_2,L^{(0)},\Theta)
 \nonumber \\
 &\hspace{55mm}
 -\frac{\partial_IG_{1}(H_2,L^{(0)},\Theta)}{\Omega}\,L^{(1)}(q_2,p_2,H,\Theta)
 \biggr] \partial_{p_2}H_2(q_2,p_2)\,,
 \nonumber \\
 F^{(0)}
 &=-\frac{G_{1}(H_2,L^{(0)},\Theta)}{\Omega}\,\partial_{q_2}H_2(q_2,p_2)\,,
 \nonumber \\
 F^{(1)}
 &=-\frac{G_{\mathrm{int},q}(q_2,p_2,L^{(0)},\Theta)}{\Omega}
 +\biggl[
 \frac{G_{1}(H_2,L^{(0)},\Theta)}{\Omega^2}\, \tilde\Omega(q_2,p_2,L^{(0)},\Theta)
 \nonumber \\
 &\hspace{55mm}
 -\frac{\partial_IG_{1}(H_2,L^{(0)},\Theta)}{\Omega}\,L^{(1)}(q_2,p_2,H,\Theta)
 \biggr] \partial_{q_2}H_2(q_2,p_2)\,. \label{reduced}
\end{align}
Now three variables: $q_2$, $p_2$ and $H$ are concerned with our analysis. 
Here there are some comments. The first is that $H$ is a slow variable compared to the other two.
The second is that $P^{(1)}$ and $F^{(1)}$ are periodic in $\Theta$ 
because the ratio of $\Omega$ and the period of $H_{\rm int}$ has been taken to be rational. 
This periodicity is required in Melnikov's method; hence the rational condition has been supposed
just below (\ref{rational})\,. Finally when $H$ satisfies appropriate conditions,
the reduced system (\ref{reduced}) exhibits a horseshoe for sufficiently small $\epsilon$\,,
as we will show below. 

\medskip 

Because the energy $H$ is not conserved, one needs to check whether  
the Poincar\'e map would map a volume element on a surface with constant energy 
to the same surface or another surface but close to the original.
Let us assume that there is $N$ where $(P^{\Theta_0}_{\epsilon})^N$ has a horseshoe.
$N$ is related to $\epsilon$, and the limit $\epsilon\to 0$ makes $1/N$ go zero.
The average change of energy by $(P^{\Theta_0}_{\epsilon})^N$ is
\begin{align}
 \mathit{\Delta}H
 &=\int_{-\pi N}^{\pi N}\!\! d\Theta \, \overline{H'}
 =\int_{-\pi N/\Omega}^{\pi N/\Omega}\!\! d\tau \, \overline{\dot H}\,,
\end{align}
where the bar stands for the average over explicit $\Theta$ with the other variables fixed.
The integral over $\Theta$ is performed by all dependence of $\Theta$ including implicit one.
It is sufficient to evaluate this on the homoclinic orbit because the region 
where a horseshoe structure appears is close to the homoclinic orbit for small $\epsilon$\,.
Now, we assume that, for a certain value of $H$\,, 
\begin{align}
 \mathit{\Delta}H =0
 \qquad
 \text{with any $N$ sufficiently large}.
 \label{DeltaH=0}
\end{align}
Then, $(P^{\Theta_0}_{\epsilon})^N$ maps the energy surface to itself 
to leading order in $\epsilon$.
This means that there exists a map from a vicinity near the saddle point to itself
under this condition\footnote{There is another type of condition that is not used in this paper.
See \cite{HolmesMarsden:1982} for further information.}.

\medskip 

Before defining the Melnikov function in this case,
one needs to remove the explicit $\Theta$-dependence 
in $P^{(0)}$ and $F^{(0)}$.
Since $H$ and $H_2$ are nearly constant as far as the Poincar\'e section is concerned,
one can expand those in $\epsilon$ as
\begin{align}
 H(\Theta)&=H^{(0)}+\epsilon H^{(1)}(\Theta)+O(\epsilon^2)\,,
\end{align}
and
\begin{align}
 H_2(\Theta)&=H_2^{(0)}+\epsilon H_2^{(1)}(q_2,p_2,\Theta)+O(\epsilon^2)\,.
\end{align}
Here, explicit $\Theta$ dependence in $H_2^{(1)}$ comes from that in the separatrix solution.
With an assumption that $G_1(H_2^{(0)},L^{(0)}(H_2^{(0)},H^{(0)}),\Theta)$ is positive-definite,
we now define a new time $T$ by
\begin{align*}
 \frac{dT}{d\Theta}=\frac{G_1(H_2^{(0)},L^{(0)}(H_2^{(0)},H^{(0)}),\Theta)}{\Omega}\,.
\end{align*}
Then, functions periodic in $\Theta$ are periodic in $T$ as well if
\begin{align}
 T_{\rm period}\equiv \int_{-\pi}^{\pi} d\Theta\, \frac{dT}{d\Theta}
 \label{Tperiod}
\end{align}
is finite since $G_1(H_2^{(0)},L^{(0)}(H_2^{(0)},H^{(0)}),\Theta)$ is periodic in $\Theta$.
One finds the equations of motion of $q_2$ and $p_2$ finally become
\begin{align}
 \frac{dq_2}{dT}
 =\mathcal{P}^{(0)}+\epsilon\, \mathcal{P}^{(1)}+O(\epsilon^2)\,,
 \qquad
 \frac{dp_2}{dT}
 =\mathcal{F}^{(0)}+\epsilon\, \mathcal{F}^{(1)}+O(\epsilon^2)\,,
 \label{eq:q,p,T}
\end{align}
where we have introduced the following quantities: 
\begin{align}
 &\mathcal{P}^{(0)}
 =\partial_{p_2}H_2(q_2,p_2)\,,
 \nonumber \\
 &\mathcal{P}^{(1)}
 =\frac{G_{\mathrm{int},p}(q_2,p_2,L^{(0)},\Theta(T))}{G_{1}(H_2^{(0)},L^{(0)},\Theta(T))}
 -\mathcal{G}(q_2,p_2,H,\Theta(T))\, \partial_{p_2}H_2(q_2,p_2)\,,
 \nonumber \\
 &\mathcal{F}^{(0)}
 =-\partial_{q_2}H_2(q_2,p_2)\,,
 \nonumber \\
 &\mathcal{F}^{(1)}
 =-\frac{G_{\mathrm{int},q}(q_2,p_2,L^{(0)},\Theta(T))}{G_{1}(H_2^{(0)},L^{(0)},\Theta(T))}
 +\mathcal{G}(q_2,p_2,H,\Theta(T))\, \partial_{q_2}H_2(q_2,p_2)\,, 
 \nonumber \\
 &\mathcal{G}(q_2,p_2,H,\Theta)
 =\frac{\Omega}{G_{1}(H_2^{(0)},L^{(0)},\Theta)}\,\biggl[ 
 \frac{G_{1}(H_2^{(0)},L^{(0)},\Theta)}{\Omega^2}\, \tilde\Omega(q_2,p_2,L^{(0)},\Theta)
 \nonumber \\
 &\hspace{65mm}
 -\frac{\partial_IG_{1}(H_2^{(0)},L^{(0)},\Theta)}{\Omega}\,
 \Big( L^{(1)}(q_2,p_2,H,\Theta)
 \nonumber \\
 &\hspace{75mm}
 +\partial_{H_2}L^{(0)}H_2^{(1)}(q_2,p_2,\Theta)+\partial_{H}L^{(0)}H^{(1)}(\Theta) \Big)
 \nonumber \\
 &\hspace{65mm}
 -\frac{\partial_{H_2}G_{1}(H_2^{(0)},L^{(0)},\Theta)}{\Omega}\,H_2^{(1)}(q_2,p_2,\Theta)
 \biggr] \,.
\end{align}

\medskip

Because this is a one-dimensional system, to which one can apply Melnikov's method,
the Melnikov function is straightforwardly defined as 
\begin{align}
 M(T_0)
 &\equiv \int_{-\infty}^\infty\!\! dT\,
 (\mathcal{P}^{(0)}\mathcal{F}^{(1)}-\mathcal{F}^{(0)}\mathcal{P}^{(1)})
 (q_{2}^{(0)}(T-T_0),p_{2}^{(0)}(T-T_0),H^{(0)},T)\,.  
 \label{mel}
\end{align}
Here $q_{2}^{(0)}$ and $p_{2}^{(0)}$ are
the separatrix solution to \eqref{eq:q,p,T} at $\epsilon=0$\,,
and $H^{(0)}$ is the initial total energy.

\medskip 

Let us rewrite (\ref{mel}) as a function of $\tau$\,, rather than $T$\,.
We define $\tau_0$ as 
\[
\tau_0 \equiv \Theta(T_0)/\Omega +O(\epsilon)
\] 
and denote the value of $H_2$ with the separatrix solution by $H_{2}^{(0)}$\,.
Then, by using the following notations:
\begin{eqnarray}
q_{1}^{(0)}(\tau) &=& q_{1}\left(L^{(0)}(H_{2}^{(0)},H^{(0)}),\Omega\tau,H_{2}^{(0)}
\right)\,, \nonumber \\
p_{1}^{(0)}(\tau) &=& p_{1}\left(L^{(0)}(H_{2}^{(0)},H^{(0)}),\Omega\tau,H_{2}^{(0)}
\right)\,, 
\end{eqnarray}
the Melnikov function can be rewritten as 
\begin{align}
 M(\tau_0)
 &=\int_{-\infty}^\infty\!\! d\tau\, 
 \{ H_2, H_{\rm int} \} (q_1^{(0)}(\tau),p_1^{(0)}(\tau),q_{2}^{(0)}(\tau-\tau_0),
p_{2}^{(0)}(\tau-\tau_0),H^{(0)},\tau) \notag \\ 
& \qquad +O(\epsilon)\,.
 \label{Melnikov fn spec}
\end{align}
This function is proportional to the separation of the stable and unstable manifolds 
at $\tau=\tau_0$ near the surface of $H=H^{(0)}$ 
because $\mathit{\Delta}H=0$ is assumed. Hence, Melnikov's method says that
the existence of simple zeros in $M(\tau_0)$ indicates that this system has a horseshoe 
for sufficiently small $\epsilon$ lying near the homoclinic orbit of $(q_2,p_2)$ 
and near the energy surface $H=H^{(0)}$\,.

\section{A brane-wave deformation of AdS$_5\times$S$^5$ \label{3:sec}}

In this section, we shall consider an application of Melnikov's method 
in the context of String Theory. In other words, we are concerned with 
an appropriate string background to which Melnikov's method is applicable.   

\medskip 

An example of such backgrounds is a brane-wave type deformation of the AdS$_5\times$S$^5$ background 
constructed in \cite{HY}. 
The background is composed of the metric and the five-form field strength  
\begin{eqnarray}
{\rm ds}^2&=& {\rm ds}^2_{\rm AdSpp} + {\rm ds}^2_{\rm S^5}\,, \nonumber \\ 
{\rm ds}^2_{\rm AdSpp} &=& L^2\left[\frac{-2{\rm d}x^+{\rm d}x^- + ({\rm d}x^1)^2+({\rm d}x^2)^2+{\rm d}z^2}{z^2} 
-\frac{\eta^2}{z^{2a}}\,f(\mu,\theta)\,
({\rm d}x^+)^2 \right]\,, \label{metric}  \\
{\rm ds}^2_{\rm S^5}  &=& {\rm ds}^2_{{\mathbb C}\rm P^2}+({\rm d}\chi+\omega)^2\,, \nonumber \\ 
F_5 &=& 4 [\omega_{\rm AdS_5} + \omega_{\rm S^5}]\,, \nonumber 
\end{eqnarray}
where $a$ satisfies $a=(l+2)/2$ with $l$ an integer greater than 0.
The dilaton is constant and the other fluxes are zero. 
Here the metric of round S$^5$ is expressed as a $U(1)$ fibration over 
${\mathbb C}$P$^2$\,, where $\chi$ is the local coordinate on the Hopf fibre and $\omega$ 
is the one-form potential for the K\"ahler form on ${\mathbb C}$P$^2$\,. 
The metric of ${\mathbb C}$P$^2$ and $\omega$ are given by  
\begin{eqnarray}
{\rm ds}^2_{{\mathbb C}\rm P^2}&=&
{\rm d}\mu^2+\sin^2\mu\left(\Sigma_1^2+\Sigma_2^2+\cos^2\mu\,\Sigma_3^2\right)\,, 
\qquad
\omega=\sin^2\mu\,\Sigma_3\,, 
\end{eqnarray}
where $\Sigma_a~(a=1,2,3)$ are defined as 
\begin{eqnarray}
\Sigma_1& \equiv &\frac{1}{2}\left(\cos\psi\, {\rm d}\theta +\sin\psi\sin\theta\, {\rm d}\phi\right)\,, 
\quad 
\Sigma_2 \equiv \frac{1}{2}\left(\sin\psi\, {\rm d}\theta -\cos\psi\sin\theta\, {\rm d}\phi\right)\,, 
\nonumber \\
\Sigma_3& \equiv&\frac{1}{2}\left({\rm d}\psi+\cos\theta\, {\rm d}\phi\right)\,. 
\nonumber
\end{eqnarray}

\medskip 

The remaining task is to fix an arbitrary function $f(\mu,\theta)$\,. 
A remarkable point is that the metric (\ref{metric}) is a pp-wave type deformation; 
hence the only non-trivial equation of motion is the $(+,+)$-component of the equations 
of motion for the metric. Thus all we have to do is to solve the following constraint: 
\begin{eqnarray}
&& 4(a^2-1) f(\mu,\theta) + \frac{4}{\sin^2 \mu} 
\left(\cot\theta \partial_{\theta}f(\mu,\theta) 
+ \partial_{\theta}^2f(\mu,\theta)\right) \nonumber \\ 
&& \hspace*{2cm}
+ \left(3\cot\mu -\tan\mu\right)\partial_{\mu}f(\mu,\theta) 
+ \partial_{\mu}^2f(\mu,\theta) =0\,. \label{cons}
\end{eqnarray}
Note here that this constraint (\ref{cons}) is obviously satisfied when $a=1$ and $f(\mu,\theta)$ is a constant. 
This case corresponds to the undeformed AdS$_5\times$S$^5$\,.   
It is easy to derive the general solution of (\ref{cons}), 
but some simple ones are enough for our purpose here. 
Hereafter, we will concentrate on two solutions given by  
\begin{eqnarray}
f(\mu,\theta) =
\left\{ 
\begin{array}{cc}
\sin^2\mu \cos\theta & \qquad \mbox{for} ~~ a=2 \\ 
(1-6\cos^2\mu+5\cos^4\mu)\cos\theta  & \qquad \mbox{for} ~~ a=3
\end{array}
\right.\,. \label{f(mu)}
\end{eqnarray}
Here the overall constant can be absorbed by rescaling $\eta$\,; 
hence we have simply set it to 1. 

\medskip

We will study the motion of a string propagating on the background presented above. 
To show classical chaos in the bosonic sector, we will not touch on the fermionic sector. 
In addition, to make our analysis simpler, we will adopt two ans\"atze. 
In the following, we will compute Melnikov's function under each of the ans\"atze.

\subsection{A reduction ansatz: an oscillating string}\label{sec:oscil_ansatz}

In this subsection, we will consider an oscillating string ansatz. 

\medskip 

Let us perform a coordinate transformation \cite{Kruczenski:2008bs}:
\begin{align}
 &x^+=\frac{1}{m}\tan \left(m \tilde x^+\right)\,,
 \qquad
 x^-=x^-+\frac{m}{2}\left[(\tilde{x}^1)^2 + (\tilde{x}^2)^2  + \tilde z^2\right]
\tan \left(m \tilde x^+\right)\,,
 \nonumber \\
 &x^i=\frac{\tilde x^i}{\cos \left(m \tilde x^+\right)} \qquad (i=1,2)\,,
 \qquad
 z=\frac{\tilde z}{\cos \left(m \tilde x^+\right)}\,. \label{singular}
\end{align}
Note here that the points with $m\tilde{x}^+ = n \pi ~~(n \in \mathbb{Z})$ are singular.
After this transformation, the system gets restricted and 
a certain part of the deformed AdS background is magnified.

\medskip

The resulting metric is given by 
\begin{eqnarray}
 {\rm ds}^2
  &=&L^2\left[
	 \frac{-2{\rm d}\tilde x^+{\rm d}\tilde x^-
	 -m^2(\tilde x^{i\; 2}+\tilde z^2)({\rm d}\tilde x^+)^2
	 +({\rm d}\tilde x^i)^2
	 +{\rm d}\tilde z^2}{\tilde z^2} 
	\right. \nonumber \\
 &&\hspace{1cm} 
  \left. 
   -\frac{\eta^2}{\tilde z^{2a}}\cos^{2a-4}(m \tilde x^+)\,f(\mu,\theta)\, ({\rm d}\tilde x^+)^2 
   + {\rm ds}^2_{\rm S^5}
  \right]\,.
\end{eqnarray}
Note here that the undeformed AdS$_5$ is also deformed to a pp-wave type metric. 

\medskip 

In the following, we are concerned with a string theory defined on this background. 
We will concentrate on the bosonic part and study the classical action of Polyakov type 
with the light-cone gauge. 

\medskip 

For simplicity, let us take the following ansatz: 
\begin{eqnarray}
 && \tilde x^+ = \tau\,, \qquad \tilde x^1=\tilde x^2 =0\,, \qquad \tilde z=Z(\tau)\,, \nonumber \\ 
 && \mu=\mu(\tau)\,, \qquad \phi = n\, \sigma\,, \qquad \theta = \chi = \psi =0\,. \label{ansatz1}
\end{eqnarray}
Here $n \in \mathbb{Z}$ is a winding number along the $\phi$-direction. 
Note that this ansatz is consistent with the original equations of motion. 

\medskip  

From now on, let us set $a=3$ for simplicity and the function $f(\mu,\theta)$ is given in (\ref{f(mu)}). 
This is just because the case with $a=2$ is 
integrable under the ansatz (\ref{ansatz1}) and hence the case with $a=3$ is the simplest and 
non-integrable, though the case with $a>3$ would be non-integrable as well.  

\medskip 

Then the light-cone Hamiltonian is simplified as\footnote{
Here we have used a formula of the light-cone Hamiltonian obtained by solving the Virasoro constraints. 
For example, see \cite{Callan:2003xr:Swanson:2005wz}.}
\begin{eqnarray}
 {\cal{H}}_{\rm lc} 
  &=&
   \frac{1}{2}\left[
	       p_{Z}^2+m^2Z^2
	       +\frac{1}{Z^2}\left( p_\mu^2
	       +\frac{n^2L^4}{4}\sin^2\mu \right)
	       +\frac{\eta^2f(\mu,0)}{Z^{4}}\cos^{2}(m\tau)
	      \right]\,.
 \label{light cone Hamiltonian g_--=0}
\end{eqnarray}
This is nothing but a coupled pendulum-oscillator system. 
It should be remarked that the Hamiltonian (\ref{light cone Hamiltonian g_--=0}) has the explicit 
$\tau$-dependence due to the singular coordinate transformation (\ref{singular})\,.  
The unperturbed Hamiltonian is
\begin{align}
 H_{1} = \frac{1}{2}p_Z^2 + \frac{1}{2}m^2Z^2 + \frac{H_{2}}{Z^2}\,, \label{lc1}
\end{align}
where $H_2$ is the Hamiltonian of a pendulum system,
\begin{align}
 H_{2} = \frac{1}{2}p_\mu^2+\frac{n^2L^4}{8}\sin^2\mu\,.
\end{align}
The perturbation term is given by 
\begin{align}
 \eta^2 H_{\rm int} = \eta^2\,\frac{f(\mu,0)}{2Z^{4}}\,\cos^{2}(m\tau)\,,
\end{align}
where $\eta$ is assumed to be infinitesimal. 
For this system, one can compute the Melnikov function by following the generalized method developed 
in Sec.~\ref{sec:spec-meln-meth}.

\medskip 

In the non-perturbed case (i.e., $\eta=0$), $H_2$ can be regarded as a constant $h$ 
because $\{\mathcal{H}_{\rm{lc}},H_2\}=0$\,.
Then the Hamiltonian can be rewritten in terms of the action-angle variables 
by performing a canonical transformation to $(I,\Theta)$~:
\begin{align}
 Z  &=\sqrt{\frac{2I+\sqrt{2h}+2\sqrt{I(\sqrt{2h}+I)}\cos \Theta}{m}}\,, 
 \nonumber \\
 p_Z &=-\sqrt{\frac{4mI(\sqrt{2h}+I)}{2I+\sqrt{2h}+2\sqrt{I(\sqrt{2h}+I)}\cos \Theta}}\,\sin\Theta \,.
\label{canonical-trf}
\end{align}
The resulting Hamiltonian is expressed in terms of $(I,\Theta)$ as follows:\footnote{The appearance of the square root 
in (\ref{sq}) is conceivable because the third term in (\ref{lc1}) can be regarded 
as a centrifugal force with a angular momentum $L_o$ under the identification $h \simeq L_o^2$\,.} 
\begin{align}
\mathcal{H}_{\rm{lc}} |_{\eta=0}=E=2m I+m \sqrt{2h}\,. \label{sq}
\end{align}
Here we have introduced the energy $E$\,. From the equation of motion of $(I,\Theta)$\,,
\begin{align}
I =\frac{1}{2}\left(\frac{E}{m}-\sqrt{2 h}\right)=\text{const.}\,,\qquad \Theta=2m (\tau-\tau_1)\,.
\end{align}
Now $Z$ is written as a function of $\tau$ through the transformation (\ref{canonical-trf})\,,
\begin{align}
Z^{(0)}(\tau)=\sqrt{\frac{E}{m^2}}\sqrt{1+\sqrt{1-\frac{2hm^2}{E^2}}\cos [2m(\tau-\tau_1)]}\,.
\end{align}

\medskip

In order to calculate the Melnikov function, the separatrix solution of the pendulum system $H_2$ 
is necessary. 
If one takes 
\begin{eqnarray}
\frac{dT}{d\tau}=\frac{1}{Z^2}\,, \label{tr}
\end{eqnarray} 
then $\mathcal{P}^{(0)}$ and $\mathcal{F}^{(0)}$ in \eqref{eq:q,p,T}, in this system, are given by 
\begin{align}
 \mathcal{P}^{(0)}&=p_\mu\,, 
 \nonumber \\
 \mathcal{F}^{(0)}&=-\frac{n^2L^4}{8}\sin2\mu \,,
\end{align}
and the equations of motion for the pendulum system with time $T$ by
\begin{align}
 \frac{d\mu}{dT}&=p_\mu \,,
 \qquad
 \frac{dp_\mu}{dT}=-\frac{n^2L^4}{8}\sin2\mu \,,
\end{align}
which can easily be solved by 
\[
\mu=2\Arctan \left[\tanh \frac{nL^2}{4}\,T \right]\,.
\]
By integrating (\ref{tr})\,, the following solution is obtained,
\[
T(\tau)=\frac{1}{\sqrt{2h}}
\Arctan\left[ \frac{E-\sqrt{E^2-2hm^2}}{\sqrt{2hm^2}}
\tan[m(\tau-\tau_1)]\right]\,.
\]
The new period \eqref{Tperiod} becomes $T_{\rm period}=\pi/\sqrt{2h}$\,.
Now, with $T(\tau)$\,,  the separatrix solution can be rewritten as
\begin{align}
 \mu^{(0)}(\tau)=2\Arctan \left[
 \tanh \frac{nL^2}{4}T(\tau)
 \right] \,.
\end{align}
Note that its energy $h$ is $h^{(0)}=n^2L^4/8$\,.

\medskip 

One must check the energy condition \eqref{DeltaH=0}\,.
The equation of the energy is 
\begin{align}
 \mathcal{\dot H}_{\rm lc} 
 &=-\eta^2\frac{f(\mu,0)}{4(Z^{(0)}(\tau))^{4}}
 \sin (2m\tau)
 +O(\eta^4)\,. 
\end{align} 
Its average on the homoclinic orbit is given by 
\begin{align}
 \overline{\mathcal{\dot H}_{\rm lc}}
 &=-\eta^2\frac{mf(\mu^{(0)},0)}{8\pi}\int_{-\pi/m}^{\pi/m}d\tau
 \frac{\sin (2m(\tau+\tau_1))}{(Z^{(0)}(\tau))^{4}}
 +O(\eta^4) \nonumber \\
 &\simeq \eta^2\frac{2m^2f(\mu^{(0)},0)}{(nL^2)^2}\, A
 \sin (2m\tau_1)\,.
\end{align}
Here $\mathcal{H}_{\rm lc}$ is an approximation in the region 
sufficiently near 
\[
m\sqrt{2h^{(0)}} = \frac{1}{2} mnL^2\,.
\]
We also use the following expression, 
\[
A\equiv \sqrt{\frac{E}{mnL^2}-\frac{1}{2}}
\,, 
\]
which is sufficiently small.

\medskip 

Then, the change of the energy by a Poincar\'e map 
with $N$ shifts of time by the period $2\pi/m$ is estimated as
\begin{align}
 \mathit{\Delta}\mathcal{H}_{\rm lc}
 &=\eta^2
 \frac{2m^2}{(nL^2)^2}\,A\,
 \sin (2m\tau_1)
 \int_{\tau_1-\frac{\pi N}{m}}^{\tau_1+\frac{\pi N}{m}} d\tau \, f(\mu^{(0)},0)
 \nonumber \\
 &\to 
 \frac{\eta^2A}{(nL^2)^2}
 \left(
 2\pi N-\frac{16}{3}
 \right) 
 \sin (2m\tau_1)\,,
 \label{DeltaH_lc}
\end{align}
at $N\to\infty$\,. In the above calculation, we have omitted the subleading order in $A$
and used an approximate relation 
\[
T \simeq \frac{2m}{nL^2}(\tau -\tau_1)\,. 
\] 
The equation \eqref{DeltaH_lc} suggests that the total energy generally changes monotonically.
Eventually, we can expect this change from the beginning 
because the frequency of the factor oscillating explicitly  
in the perturbed Hamiltonian coincides with the frequency of unperturbed $Z$\,.
Hence, it possibly provides a ``resonance'' source.
The condition \eqref{DeltaH=0} is however satisfied when 
\[
\tau_1=\frac{\pi}{2m}l \qquad (l\in \mathbb{Z})\,.
\]
If a string with this condition starts its trajectory from the vicinity near the homoclinic point,
then the energy will not change drastically 
when the string comes back to the vicinity near the homoclinic point.

\medskip

Now that we can apply the Melnikov's method explained in Sec.~\ref{sec:spec-meln-meth}\,, 
the Melnikov function \eqref{Melnikov fn spec} is computed as 
\begin{align}
 M(\tau_0)
 =\int_{-\infty}^\infty d\tau\, 
 \{ H_{2}^{(0)},H_{\rm int} \} (\tau-\tau_0)\,.
\end{align}
Here $H_{2}^{(0)}$ is defined as $H_{2}^{(0)}\equiv H_{2}(\mu^{(0)},p^{(0)}_\mu)$\,. 
This Poisson bracket is reduced to
\begin{align}
 &\{ H_{2}^{(0)},H_{\rm int} \} (\tau-\tau_0)
 \nonumber \\
 &=-\frac{1}{2}\frac{d\mu^{(0)}}{dT}\sin \left[ 2\mu^{(0)}(\tau-\tau_0) \right] 
 (1-5\cos[2\mu^{(0)}(\tau-\tau_0)] )
 (Z^{(0)}(\tau))^{-4}
 \cos^{2}(m\tau)\,.
\end{align}
Then the Melnikov function is calculated as
\begin{align}
 M(\tau_0)
 &=-\frac{nL^2}{4}\int_{-\infty}^\infty \frac{d\tau}{(Z^{(0)}(\tau))^2}\, 
 \frac{\sin 2\mu^{(0)}(\tau)}{\cosh[\frac{nL^2}{2}T(\tau)]}
 (1-5\cos[2\mu^{(0)}(\tau)] )
 \nonumber \\
 &\hspace{35mm} \times 
 (Z^{(0)}(\tau))^2(Z^{(0)}(\tau+\tau_0))^{-4}
 \cos^{2}[m(\tau+\tau_0)]
 \nonumber \\
 &= -nL^2\int_{-\infty}^\infty dT\, 
 \left(
 3\frac{\sinh[\frac{nL^2}{2}T]}{\cosh^{3}[\frac{nL^2}{2}T]}
 -5\frac{\sinh[\frac{nL^2}{2}T]}{\cosh^{5}[\frac{nL^2}{2}T]}
 \right)
 \nonumber \\
 &\hspace{35mm} \times 
 (Z^{(0)}(\tau(T)))^2(Z^{(0)}(\tau(T)+\tau_0))^{-4}
 \cos^{2}[m(\tau(T)+\tau_0)]
 \nonumber \\
 &=-\frac{4\pi m}{3nL^2\sinh\pi}
 \Biggl[
 \sin[2m(\tau_0+\tau_1)]
 \nonumber \\
 &\qquad
 -4A \biggl(
 \sin(2m\tau_0)
 +\frac{16}{\cosh\pi}(\sin[2m(2\tau_0+\tau_1)]-\sin[2m(\tau_0+\tau_1)])
 \biggr)\Biggr]
 \nonumber \\
 &\qquad
 +O(\eta^2,A^2)\,.
\end{align}
To derive the above expression, we have used the following approximation of $T$ and $\tau$
\begin{align}
 \tau \simeq \tau_1+\frac{nL^2}{2m}T+\frac{A}{m}\sin (nL^2T) +O(A^2)\,.
\end{align}
This Melnikov function has simple zeros at least when $A$ is sufficiently small for any $\tau_1$.
Thus the result indicates the existence of chaos generated by a Smale's horseshoe 
for trajectories satisfying $\mathit{\Delta}\mathcal{H}_{\rm lc}=0$\,, that is, 
\[
\tau_1=\frac{\pi}{2m}l \qquad (l\in\mathbb{Z})\,.
\]
Although such trajectories are quite rare,
trajectories with $\mathit{\Delta}\mathcal{H}_{\rm lc}\sim 0$ remain 
in a small range of the energy after performing each of the Poincar\'e maps
if the observation time is taken to be sufficiently short.

\subsection{A reduction ansatz: a spinning string}\label{sec:spin_ansatz}

In this subsection, we consider a spinning string ansatz.

\medskip

In order to obtain a desired ansatz,
we convert the Poincar\'e coordinate of the AdS part 
to the global coordinate by performing the following transformation: 
\begin{align}
 \frac{1}{z}
 &=L(\cosh\rho\sin t+\sinh\rho\cos\xi\cos\eta)\,,
 \nonumber \\
 \frac{L}{z}x^{\pm}
 &=\frac{L}{\sqrt{2}}(\cosh\rho\cos t\pm \sinh\rho\cos\xi\sin\eta)\,,
 \nonumber \\
 \frac{L}{z}x^{1}
 &=L\sinh\rho\sin\xi\sin\zeta\,,
 \nonumber \\
 \frac{L}{z}x^{2}
 &=L\sinh\rho\sin\xi\cos\zeta\,.
\end{align}

\medskip

As in the previous subsection, 
we will study the bosonic part of the Polyakov action. 
As a remark, we will take here the spatial direction of the string-world-sheet to be infinite 
(i.e., an infinitely long string), rather than the usual closed string. 
With the gauge condition $t=\kappa\tau$\,, let us consider the following ansatz: 
\begin{eqnarray}
&& t = \kappa \tau\,, \quad 
 \rho=\rho(\sigma)\,, \quad 
 \xi=\frac{\pi}{2}\,, \quad
 \eta=-\kappa \tau\,, \quad
 \zeta=\omega_1\tau \,, \nonumber \\ 
&& \mu=\mu(\sigma)\,, \quad 
 \phi = \omega_2\tau\,, \quad 
 \theta = \chi = \psi =0\,. \label{ansatz2}
\end{eqnarray}
Because we are considering an infinite string,  
there is no need to impose a boundary condition on $\mu(\sigma)$\,.
The relevant part of the metric is rewritten as 
\begin{align}
 {\rm ds}^2
 =& L^2\left[
 -\cosh^2\rho\, {\rm d}t^2+ {\rm d}\rho^2+\sinh^2\rho\, {\rm d}\zeta^2
 -\eta^2\frac{L^{2a-2}}{2}f(\mu,0)\cosh^{2a}\rho\, \sin^{2a-4}t\, {\rm d}t^2 \right. \notag \\ 
 & \left. \qquad + {\rm d}\mu^2+\frac{1}{4}\sin^2\mu\, {\rm d}\phi^2
 \right]\,.
\end{align}

\medskip 

For simplicity, we concentrate on the case with $a=2$ hereafter. 
The function $f(\mu,\theta)$ is given in (\ref{f(mu)})\,. 
Note here that this case is not integrable under the ansatz (\ref{ansatz2}) 
in comparison to the previous ansatz (\ref{ansatz1})\,.

\medskip

Then the Lagrangian density is given by 
\begin{align}
 \mathcal{L}
 =-\frac{1}{2}\left[
 \kappa^2\cosh\rho^2+\rho'^2-\omega_1^2\sinh^2\rho
 +\eta^2\frac{\kappa^2}{2}f(\mu,0)\cosh^4\rho
 +\mu'^2-\frac{\omega_2^2}{4}\sin^2\mu
 \right] \,,
 \label{lagrangian}
\end{align}
where the prime here is defined as the derivative with respect to $\sigma$\,.
$L$ has been absorbed into $\kappa$\,, $\sigma$\,, $\omega_i$\,, and $\eta$
so that $\rho$ and $\mu$ are canonically normalized.
Also, we impose a condition on $\kappa$ and $\omega_1$ like  
\begin{eqnarray}
\omega_1^2>\kappa^2 
\label{kappa}
\end{eqnarray}
so that the system should be bounded. 

\medskip 

If one regards $\sigma$ as time, 
``Hamiltonian,'' which governs the $\sigma$-dependence of $\rho$ and $\mu$\,,
can be defined as 
\begin{align}
 H \equiv \frac{1}{2}\left(
 p_\rho^2+(\omega_1^2-\kappa^2)\sinh^2\rho
 +p_\mu^2+\frac{\omega_2^2}{4}\sin^2\mu
 -\eta^2\frac{\kappa^2}{2}f(\mu,0)\cosh^4\rho
 \right)\,.
 \label{hamiltonian}
\end{align}
Thanks to the condition (\ref{kappa})\,, the second term is positively definite. 
The classical solution should satisfy the Virasoro constraints,  
\begin{align}
 0=-\kappa^2\cosh\rho^2+\rho'^2+\omega_1^2\sinh^2\rho
 +\mu'^2+\frac{\omega_2^2}{4}\sin^2\mu
 -\eta^2\frac{\kappa^2}{2}f(\mu,0)\cosh^4\rho\,. \label{Vir}
\end{align}
Note that this condition (\ref{Vir}) is exactly the same as the equation for a Noether charge for the translation 
along the $\sigma$-direction in \eqref{lagrangian} where the charge is $\kappa^2/2$\,.
Thus there is not any contradiction as far as the total ``energy'' is $\kappa^2/2$\,.
Moreover, as far as the Virasoro constraints are satisfied, 
the ansatz \eqref{ansatz2} is consistent with the original equations of motion. 
In the following, we treat $\sigma$ like time and use the symbol ``$\tau$'', instead of $\sigma$\,. 

\medskip

This system is regarded as a coupled pendulum-oscillator system.
The Hamiltonian of an oscillator system is given by 
\begin{align}
 H_{\rm osc}=\frac{1}{2}\left(
 p_\rho^2+(\omega_1^2-\kappa^2)\sinh^2\rho
 \right)\,,
\end{align}
and that of a pendulum system is
\begin{align}
 H_{\rm pen}=\frac{1}{2}\left(
 p_\mu^2+\frac{\omega_2^2}{4}\sin^2\mu
 \right)\,.
\end{align}
While the latter system is a usual pendulum one,
the former is oscillated in a squared hyperbolic sine potential.
The interaction term is written as
\begin{align}
 \eta^2 H_{\rm int}=
 -\eta^2\frac{\kappa^2}{4}f(\mu,0)\cosh^4\rho\,,
\end{align}
where $\eta$ is assumed to be infinitesimal. 
When $\eta=0$\,, the Hamiltonian system is a direct sum of 
two systems where one is an oscillator and the other has a homoclinic orbit.
Hence to compute the Melnikov function, we can follow the standard method described 
in Sec.~\ref{sec:stand-meln-meth}.

\medskip

The solution of the oscillator system with its energy $E$ is
\begin{align}
 \rho(\tau)=-i\operatorname{am}
 \left( i\sqrt{2E}\,\tau,i\sqrt{\frac{\omega_1^2-\kappa^2}{2E}} \right)
 \simeq \sqrt{\frac{2E}{\omega_1^2-\kappa^2}}\,
 \sin\left[ \sqrt{\omega_1^2-\kappa^2}\,\tau \right]\,,
 \label{solution_osc}
\end{align}
where the function am$(x,k)$ is the Jacobi amplitude function, 
defined as the inverse of the elliptic integral of the first kind $F(x,k)$ as
\begin{align}
 x=F\left(\operatorname{am}(x,k),k\right),
 \qquad
 F(x,k)=\int_0^x\frac{dy}{\sqrt{1-k^2\sin^2y}}\,.
 \label{Jacobi-amp}
\end{align}
The last expression in \eqref{solution_osc} is obtained as an approximation 
when $E$ is sufficiently small.

\medskip

Let us compute the Melnikov function by using the separatrix solution of the pendulum system, 
\begin{align}
 \mu^{(0)}(\tau)=2\Arctan \left[
 \tanh \frac{\omega_2(\tau-\tau_0)}{4}
 \right]\,. 
\end{align}
The energy of this solution is $\omega_2^2/8$\,.
As the total energy is $\kappa^2/2$\,,
the value of $E$ is estimated as $(4\kappa^2-\omega_2^2)/8$\,, if the interaction term is negligible.

\medskip

The present case does not require the energy condition 
because the system is not disturbed by an external force. 
Hence the Melnikov function defined in \eqref{Melnikov fn reduced} 
is given by 
\begin{align}
 M(\tau_0)
 =\int_{-\infty}^\infty\!\! d\tau\, \{ H_{\rm pen}^{(0)},H_{\rm int} \} (\tau-\tau_0)\,,
\end{align}
where $H_{\rm pen}^{(0)}$ is defined as $H_{\rm pen}^{(0)}\equiv H_{\rm pen}(\mu^{(0)},p_\mu^{(0)})$\,.
The Poisson bracket is now reduced to
\begin{align}
 \{ H_{\rm pen}^{(0)},H_{\rm int} \} (\tau-\tau_0)
 &=\frac{\kappa^2}{4}\dot\mu^{(0)}(\tau)\sin \left( 2\mu^{(0)}(\tau) \right)
 \nd^4\left( \omega'_1\tau,s \right)\,, \\ 
 \omega'_1 &=\sqrt{2E+\omega_1^2-\kappa^2}\,, \qquad 
 s=\sqrt{2E/\omega'^2_1} < 1\,,  \notag 
\end{align} 
where we have used the following identities
\footnote{For the definition of elliptic functions, see Appendix \ref{elliptic functions}.}:  
\begin{align*}
 \cos[\operatorname{am}(ix,ik)]
 =\cn(ix,ik)
 =\nd\left( \sqrt{1+k^2}\, x,\frac{1}{\sqrt{1+k^2}} \right)\,.
\end{align*}
Note here that due to the condition (\ref{kappa})\,, $\omega_1$ is real and $s<1$\,.

\medskip 

The final step is to perform the integration with respect to $\tau$\,. 
It is really complicated and messy; hence the detailed calculation is explained in Appendix \ref{apx:calc}. 
After a lengthy calculation, the Melnikov function is given by 
\begin{align}
 M(\tau_0)
 &=\frac{\pi\kappa^2}{2}\, \sum_{k=0}^\infty \Bigg\{ 
 \frac{32E}{\omega'^2_1\omega_2^2}
 \nd\left( \omega'_1\tau_0,s \right) 
 \sd\left( \omega'_1\tau_0,s \right) 
 \cd\left( \omega'_1\tau_0,s \right) 
 g_1''\left( (2k+1)\frac{\pi}{\omega_2}\right)
 \nonumber \\
 &\hspace{30mm}
 -\sum_{l=-\infty}^{\infty}
 \operatorname{Im}\left[
 g_2\left( \tau_0-\frac{((2l+1)K+i(2k+1)K')}{\omega'_1}\right)
 \right]
 \Bigg\}\,,
 \label{Melnikov_spinning}
\end{align}
where we have introduced the following quantities 
\[
s' \equiv \sqrt{(\omega_1^2-\kappa^2)/\omega'^2_1} < 1\,,  \qquad 
K\equiv K(s)=F(\pi/2,s)\,,  \qquad K' \equiv K(s')=F(\pi/2,s')\,, 
\]
and the functions $g_1$ and $g_2$ are defined as, respectively, 
\begin{align}
 g_1(\tau)
 &\equiv \frac{-
 \nd\left( \omega'_1\tau,s' \right) 
 \sd\left( \omega'_1\tau,s' \right) 
 \cd\left( \omega'_1\tau,s' \right) }
 {\left[ 1-\frac{2E(\omega_1^2-\kappa^2)}{\omega'^4_1}
 \sd^2\left( \omega'_1\tau_0,s \right)
 \sd^2\left( \omega'_1\tau,s' \right) \right]^4} 
 \biggl[
 \nd^2\left( \omega'_1\tau_0,s \right)
 \cd^2\left( \omega'_1\tau,s' \right)
 \nonumber \\
 &\qquad\hspace{20mm}
 -\left(\frac{2E}{\omega'^2_1}\right)^2
 \sd^2\left( \omega'_1\tau_0,s \right)
 \cd^2\left( \omega'_1\tau_0,s \right)
 \sd^2\left( \omega'_1\tau,s' \right)
 \nd^2\left( \omega'_1\tau,s' \right)
 \biggr]\,,
 \nonumber \\
 g_2(\tau)
 &\equiv \frac{-\omega_2^2}
 {24\left( 1-\frac{2E}{\omega'^2_1}\right)^2\omega'^4_1\cosh^6(\frac{\omega_2\tau}{2})}
 \biggl[
 16(\omega'^2_1-E+\omega_2^2)
 +(8(\omega'^2_1-E)-13\omega_2^2)\cosh(\omega_2\tau)
 \nonumber \\
 &\qquad\hspace{50mm}
 +(-8(\omega'^2_1-E)+\omega_2^2)\cosh^2(\omega_2\tau)
 \biggr]\,. \notag
\end{align} 
The factor $\sd(\omega'_1\tau_0)\cd(\omega'_1\tau_0)$ in the first term of 
\eqref{Melnikov_spinning} has non-trivial zeros periodically at $\tau_0=K l/\omega'_1$\,, 
where $l$ is integer. Then $g_1''((2k+1)\pi /\omega_2)$ also depends on $\tau_0$ 
and has a non-zero value at general $\tau_0$\,. 
Although Im$[g_2(\tau_0+((2l+1)K+i(2k+1)K')/\omega'_1)]$ in the second term
is not periodic in $\tau_0$\,, its absolute value decreases exponentially as $\tau_0$ 
goes to infinity. Then it gets smaller than the maximum of the absolute value of 
the first term when $\tau_0$ is sufficiently large.
Furthermore, when $E$ is small enough, the second term is negligible, 
and it is easier to see simple zeros as explained in Appendix \ref{apx:calc}.
Thus the Melnikov function has simple zeros at sufficiently large $\tau_0$\,,
and hence this system exhibits classical chaos associated with Smale's horseshoe.

\section{Conclusion and discussion} 

In this paper, we have considered an application of Melnikov's method in the context of 
the gauge/gravity correspondence. In particular, we have presented a string background 
to which Melnikov's method is applicable. This is a brane-wave type deformation of 
the AdS$_5\times$S$^5$ background. By employing two reduction ans\"atze, 
we have studied two types of coupled pendulum-oscillator systems. 
Then the Melnikov function has been computed for each of the systems 
and the existence of chaos has been shown analytically. 

\medskip 

A strong advantage of Melnikov's method is that the existence of classical chaos 
can be shown analytically and the mechanism is explained as a Smale's horseshoe. 
The Melnikov functions here have been computed on the string-theory side. 
One of the most interesting question is what is the gauge-theory interpretation 
of the Melnikov function. This function knows all about the classical chaotic string solutions, 
including the associated fractal structure. According to the standard AdS/CFT dictionary, 
classical string solutions should correspond to composite operators on the gauge-theory side. 
The information on the fractal structure would be crucial in determining the alignment of 
component fields in the associated composite operators. Hence the Melnikov function 
may play an important role in determining the associated composite operators in some manner. 

\medskip 

From more technical aspects, it is significant to consider a generalization to include the friction. 
Recall that an external force was induced with the oscillating ansatz. If the friction could be introduced as well, 
one can realize the system with the energy injection and dissipation. Such a system is not a conserved 
system but dissipative. The chaos in the conserved systems is well described by the KAM theorem \cite{Ko,Ar,Mo}
and its characteristic is profoundly understood. The chaos in the dissipative systems, however, 
has a richer structure such as strange attractors. It is really intriguing to realize strange attractors  
on the string-theory side and consider the physical interpretation of its gauge-theory counter part. 

\medskip 

It would also be nice to look for other string backgrounds to which Melnikov's method is applicable. 
We have just presented one example here. It is interesting to consider some applications 
of Melnikov's method for black hole solutions in relation to the information loss process. 
It may be a good issue to clarify the relation between the event horizon and the Melnikov function. 

\medskip 

We hope that Melnikov's function would provide a new tool in studying the gauge/gravity correspondence.

\subsection*{Acknowledgments}

We are very grateful to Daisuke Kawai and Takeshi Matsumoto for useful discussions. 
The work of Y.A.\ was supported by a Dublin Institute for Advanced Studies scholarship.
The work of H.K.\ was supported by the Japan Society for the Promotion of Science (JSPS).
The work of K.Y.\ was supported by the Supporting Program for Interaction-based Initiative Team Studies 
(SPIRITS) from Kyoto University and by a JSPS Grant-in-Aid for Scientific Research (C) No.\,15K05051.
This work was also supported in part by the JSPS Japan-Russia Research Cooperative Program 
and the JSPS Japan-Hungary Research Cooperative Program.

\appendix

\section*{Appendix}

\section{Elliptic functions \label{elliptic functions}}

We shall summarize the definitions of elliptic functions and useful identities utilized in Sec.~\ref{3:sec}.
Firstly, Jacobian elliptic functions are defined as follows:
\begin{eqnarray}
\sn(x,k)&\equiv& \sin [\operatorname{am}(x,k)]\,,\quad \cn(x,k)\equiv \cos [\operatorname{am}(x,k)]\,,\nonumber\\
\dn (x,k)&\equiv& \sqrt{1- k^2 \sn^2 (x,k)}\,,
\end{eqnarray}
where $\operatorname{am}(x,k)$ is the Jacobi amplitude function defined in (\ref{Jacobi-amp}). Furthermore, 
we define functions additionally as below:
\begin{eqnarray}
\sd(x,k)&\equiv& \frac{\sn (x,k)}{\dn (x,k)}\,,\quad \cd(x,k)\equiv \frac{\cn (x,k)}{\dn (x,k)}\,,\nonumber\\
\nd (x,k)&\equiv& \frac{1}{\dn (x,k)}\,.
\end{eqnarray}
There are addition theorems for the Jacobian elliptic functions. For example,
$\nd(x,k)$ satisfies the following equation: 
\begin{eqnarray}
 \nd (x+y,k) =\frac{\nd(x,k)\nd(y,k)+k^2 \sd(x,k)\sd(y,k)\cd(x,k)\cd(y,k)}{1+k^2k'^2 \sd^2(x,k)\sd^2(y,k)}\,.
 \label{addition thm}
\end{eqnarray}
Lastly, we present useful identities of Jacobian elliptic functions,
\begin{eqnarray}
\cn(ix,k)=\frac{1}{\cn(x,k')}\,,\quad \sn(x,k)=\frac{1}{k}\sn(kx,\frac{1}{k})\,,
\end{eqnarray}
where $k'=\sqrt{1-k^2}$\,.
One can show these identities by using the integral expressions of the elliptic functions.

\section{Melnikov function with a spinning string ansatz}\label{apx:calc}

We shall present here the calculation of the Melnikov function \eqref{Melnikov_spinning} in detail.

\medskip 

Let us first write down the expressions of $\dot\mu^{(0)}$ and $\sin\left(2\mu^{(0)}\right)$ like
\begin{align}
 \dot\mu^{(0)}(\tau)
 &=\frac{\omega_2}{2\cosh\frac{\omega_2(\tau-\tau_0)}{2}}\,,
 \nonumber \\
 \sin\left(2\mu^{(0)}(\tau) \right)
 &=\sin\left(4\Arctan \left[
 \tanh \frac{\omega_2(\tau-\tau_0)}{4}
 \right]\right) 
 =\frac{2\sinh\frac{\omega_2(\tau-\tau_0)}{2}}{\cosh^2\frac{\omega_2(\tau-\tau_0)}{2}}\,.
\end{align}
Then the Melnikov function is given by 
\begin{align}
 M(\tau_0) &=\frac{\kappa^2\omega_2}{4}\int_{-\infty}^\infty\!\! d\tau\, 
 \mathcal{A}(\omega_2\tau)\,
 \mathcal{B}(\omega'_1(\tau+\tau_0))\,,
 \label{Melnikov_spinning_apx}
\end{align}
where we have introduced the following new quantities: 
\begin{align}
\mathcal{A}(x) \equiv \frac{\sinh \frac{x}{2}}{\cosh^3\frac{x}{2}}\,,
\qquad \mathcal{B}(x) \equiv \nd^4\left( x, s \right) \,.
\end{align}
The integration in (\ref{Melnikov_spinning_apx}) can be evaluated 
by summing up the contributions coming from the poles of $\mathcal{A}(\omega_2\tau)$ 
and $\mathcal{B}(\omega'_1(\tau+\tau_0))$\,. In the following, let us evaluate the poles 
from $\mathcal{A}$ and $\mathcal{B}$\,, respectively.

\subsubsection*{1) the contributions coming from the poles of $\mathcal{A}$}

First of all, we shall evaluate the contributions coming from the poles of $\mathcal{A}$\,. 

\medskip 

Note that $\mathcal{A}(x)$ has triple poles at $x=(2k+1)\,\pi i~~(k\in\mathbb{Z})$\,,
and its coefficient is $-8$\,. Hence, to compute the residue, 
one needs to evaluate the second-order derivative of $\mathcal{B}(x)$\,. 

\medskip

Because $\mathcal{A}(\omega_2\tau)$ is an odd function 
and the integration range runs from $-\infty$ to $\infty$\,, 
only the odd part of $\mathcal{B}(\omega'_1(\tau+\tau_0))$ can contribute to the integral. 
Hence the calculation of derivatives of $\mathcal{B}(x)$ is simplified to some extent. 
In addition, by employing a relation \eqref{addition thm} of the Jacobian elliptic function $\nd$\,, 
one finds that the odd part of $\mathcal{B}(x+x_0)$ is given by 
\begin{align}
 &\frac{4s^2
 \nd(x_0)\sd(x_0)\cd(x_0)
 \nd(x)\sd(x)\cd(x)}
 {(1+s^2s'^2\sd(x_0)\sd(x))^4}
 \big(
 \nd(x_0)\nd(x)
 +s^4\sd(x_0)\cd(x_0)
 \sd(x)\cd(x)
 \big) \,. \notag
\end{align}
Note here that all of the second arguments of the elliptic functions are given by $s$\, 
(but those are omitted just for simplicity).

\subsubsection*{2) the contributions coming from the poles of $\mathcal{B}$}

The next task is to evaluate the contributions coming from the poles of $\mathcal{B}$\,. 

\medskip 

The Jacobian elliptic function $\nd(x,s)$ has two single poles inside the fundamental unit cell.
Two periods of $\nd(x,s)$ are given by $K $ and $4iK'$\,, where $K = K(s)$ and $K'=K(s')$\,.   
Note here that $s<1$ and $s'<1$ by construction in the present case; 
hence $K$ and $K'$ are real numbers.

\medskip 

Depending on the value of the residue, there are two kinds of poles. 
The first residue is given by $-i/s'$ for the poles at 
\[
x_{1}^{(l,k)} \equiv (2l+1)K+i(4k+1)K' \qquad (k,l\in\mathbb{Z})\,. 
\]  
The second residue is $+i/s'$ for the poles at 
\[
x_{2}^{(l,k)}  \equiv (2l+1)K+i(4k+3)K' \qquad (k,l\in\mathbb{Z})\,. 
\]  
Hence the asymptotic behaviors near these poles are given by 
\begin{align}
 \nd(x,s)
 &=\frac{-i}{s'(x-x_1^{(l,k)})}
 -\frac{i(s^2-2)}{6s'}(x-x_1^{(l,k)})
 +O((x-x_1^{(l,k)})^3) &
 \text{around}~~  x=x_1^{(l,k)}\,,
 \nonumber \\
 \nd(x,s)
 &=-\left(
 \frac{-i}{s'(x-x_2^{(l,k)})}
 -\frac{i(s^2-2)}{6s'}(x-x_2^{(l,k)})
 +O((x-x_2^{(l,k)})^3)
 \right) &  \text{around}~~  x=x_2^{(l,k)}\,. \notag
\end{align}
Now that $\mathcal{B}(x)$ has quadruple and double poles at $x=(2l+1)K+i(2k+1)K'$\,,
it is necessary to compute the third-order and first-order derivatives of $\mathcal{A}(x)$ 
in order to evaluate the residue around these poles.

\subsubsection*{Doing the integration}

Let perform the $\tau$-integration in \eqref{Melnikov_spinning_apx}. 
To perform this integral, we need to consider a contour integral. Generally, the contour depends on the behavior 
of the integrand. The contour is closed on the upper-half plane
with a semicircle in the counterclockwise direction when the integrand of \eqref{Melnikov_spinning_apx} 
goes to zero in the limit $\tau\to\infty$\,. 
Similarly, the contour should be closed on the lower-half plane with a semicircle in the clockwise direction 
depending on the behavior of the integrand in the limit $\tau\to \infty$\,. 
By denoting these contours by $C_u$ and $C_l$\,, respectively, 
the Melnikov function \eqref{Melnikov_spinning_apx} can be rewritten as
\begin{align}
 M(\tau_0)
 &=\frac{\kappa^2\omega_2}{8}\left( \int_{C_u}+\int_{C_l} \right) d\tau\, 
 \mathcal{A}(\omega_2\tau)
 \mathcal{B}(\omega'_1(\tau+\tau_0)) \,.
\end{align}

\medskip

The integration along the upper contour $C_u$ is composed of two contributions: 
1) the contributions from the poles of  $\mathcal{A}(\omega_2\tau)$\,,
namely, 
\begin{align*}
 \omega_2\tau=(2k+1)\pi i~~(k\ge 0)\,, 
\end{align*}
and 2) the contributions from the poles of $\mathcal{B}(\omega'_1(\tau+\tau_0))$\,, 
namely, 
\begin{align*}
 \omega'_1(\tau+\tau_0)=(2l+1)K+i(2k+1)K'~~(k\ge 0)\,.
\end{align*}
The former is given by 
\begin{align}
 \frac{\pi\kappa^2}{4}\,
 \frac{32E}{\omega'^2_1\omega_2^2}
 \nd\left( \omega'_1\tau_0,s \right) 
 \sd\left( \omega'_1\tau_0,s \right) 
 \cd\left( \omega'_1\tau_0,s \right) 
 \sum_{k=0}^\infty
 g_1''\left( (2k+1)\frac{\pi}{\omega_2}\right)
 \,,
 \label{residue_from_cosh^-3}
\end{align}
and the latter is 
\begin{align}
 \frac{\pi i \kappa^2}{4}\, \sum_{k=0}^\infty 
 \sum_{l=-\infty}^{\infty}
 g_2\left( \tau_0-\frac{((2l+1)K+i(2k+1)K')}{\omega'_1}\right)
 \,.
 \label{residue_from_nd^4}
\end{align}
Similarly, one can evaluate the contributions coming from the lower-half plane. 
Its contributions from the poles of  $\mathcal{A}(\omega_2\tau)$
and $\mathcal{B}(\omega'_1(\tau+\tau_0))$ are
$\omega_2\tau=-(2k+1)\pi i~~(k\ge 0)$
and $\omega'_1(\tau+\tau_0)=(2l+1)K-i(2k+1)K'~~(k\ge 0)$\,,
respectively.
Thus, the integration along the lower contour $C_l$ is evaluated as 
the same as \eqref{residue_from_cosh^-3} for the former contribution
thanks to the oddness of function $g_1$,
and the complex conjugate of \eqref{residue_from_nd^4}
for the latter contribution, respectively.
Then after summing up the integration along $C_u$ and $C_l$\,,
one finds that \eqref{Melnikov_spinning} has finally been derived.

\subsubsection*{Approximation works well} 

When $E$ is sufficiently small, one can readily derive an approximate form of 
the Melnikov function. The leading term is given by  
\begin{eqnarray}
 M(\tau_0)
 =\frac{\kappa^2\omega_2}{4}\int_{-\infty}^\infty\!\! d\tau\, 
 \frac{\sinh\frac{\omega_2\tau}{2}}{\cosh^3\frac{\omega_2\tau}{2}}
 \left( 1+\frac{4E}{\omega_1^2-\kappa^2}
 \sin^2\left[ \sqrt{\omega_1^2-\kappa^2}(\tau+\tau_0) \right] 
 \right)
 +O(E^2)\,.
\end{eqnarray}
Now that the $\tau$-integration can be performed easily, 
the resulting expression is given by 
\begin{align}
 M(\tau_0) &=
 \frac{8\pi \kappa^2E
 \sin\left[2\sqrt{\omega_1^2-\kappa^2}\tau_0\right]}{
 \omega_2^2
 \sinh\frac{2\pi\sqrt{\omega_1^2-\kappa^2}}{\omega_2}}
 +O(E^2)\,.
\end{align}
Then it is obvious that $M(\tau_0)$ has non-trivial simple zeros at 
\[
\tau_0=\frac{\pi}{2\sqrt{\omega_1^2-\kappa^2}}\,l\,.
\]
Thus the conclusion that classical chaos exists does not change even with a small energy approximation. 

\medskip 

Note here that the approximation is made for the integrand before performing the $\tau$-integration. 
One can also reproduce the same expression by expanding the exact formula \eqref{Melnikov_spinning} 
in terms of small $E$\,. That is, the order of the $\tau$-integration and the small $E$ expansion 
is irrelevant to the final result.

\end{document}